\documentclass{emulateapj}
\usepackage{apjfonts}
\usepackage{graphicx}
\usepackage{amsmath}
     \usepackage{dpfloat}

\usepackage{longtable}

\usepackage{color}

\newcommand\na{\ref@jnl{New A}}

\def\kms{\ifmmode{\rm km\thinspace s^{-1}}\else km\thinspace s$^{-1}$\fi}

\shortauthors{Sanchis-Ojeda et al.~2015}
\shorttitle{K2-22b}

\begin{document}

%
\def\ltsima{$\; \buildrel < \over \sim \;$}
\def\lsim{\lower.5ex\hbox{\ltsima}}
\def\gtsima{$\; \buildrel > \over \sim \;$}
\def\gsim{\lower.5ex\hbox{\gtsima}}
\def\teff{$T_\mathrm{eff}$}                 
 \def\vsini{\hbox{$v$\,sin\,$i_{\star}$}}    
  \def\cms2{\hbox{\,cm\,s$^{-2}$}}           

%

\def\kms{\ifmmode{\rm km\thinspace s^{-1}}\else km\thinspace s$^{-1}$\fi}

\bibliographystyle{apj}

\title{The K2-ESPRINT Project. I. Discovery of the Disintegrating Rocky Planet K2-22b \\ with a Cometary Head and Leading Tail.}

\author{
R.~Sanchis-Ojeda\altaffilmark{1, 2},
S.~Rappaport\altaffilmark{3}, 
E.~Pall\`e\altaffilmark{4, 5}, 
L.~Delrez\altaffilmark{6}, 
J.~DeVore\altaffilmark{7}, 
D.~Gandolfi\altaffilmark{8,9}, \\ 
A.~Fukui\altaffilmark{10},
I.~Ribas\altaffilmark{11}, 
K.~G.~Stassun\altaffilmark{12,13},
S.~Albrecht\altaffilmark{14},
F.~Dai\altaffilmark{3},
E.~Gaidos\altaffilmark{15},
M.~Gillon\altaffilmark{5}, \\
T.~Hirano\altaffilmark{16}, 
M.~Holman\altaffilmark{17},
A.~W.~Howard\altaffilmark{18}, 
H.~Isaacson\altaffilmark{1}, 
E.~Jehin\altaffilmark{6}, 
M. Kuzuhara\altaffilmark{16}, \\
A.~W.~Mann\altaffilmark{19, 20},
G.~W.~Marcy\altaffilmark{1}, 
P.~A.~Miles-P\'aez\altaffilmark{4, 5}, 
P.~Monta\~n\'es-Rodr\'{\i}guez\altaffilmark{4, 5},  
F.~Murgas\altaffilmark{21, 22}, \\ 
N.~Narita\altaffilmark{23, 24, 25}, 
G.~Nowak\altaffilmark{4, 5}, 
M.~Onitsuka\altaffilmark{23, 24},
M.~Paegert\altaffilmark{12},  
V.~Van~Eylen\altaffilmark{14},
J.~N.~Winn\altaffilmark{3},
L.~Yu\altaffilmark{3}.
}

\altaffiltext{1}{Department of Astronomy, University of California, Berkeley, CA 94720; sanchisojeda@berkeley.edu}

\altaffiltext{2}{NASA Sagan Fellow}

\altaffiltext{3}{Department of Physics, and Kavli Institute for
  Astrophysics and Space Research, Massachusetts Institute of
  Technology, Cambridge, MA 02139, USA, sar@mit.edu}

\altaffiltext{4}{Instituto de Astrof\'isica de Canarias (IAC), 38205 La Laguna, Tenerife, Spain}

\altaffiltext{5}{Departamento de Astrof\'isica, Universidad de La Laguna (ULL), 38206 La Laguna, Tenerife, Spain}

\altaffiltext{6}{Institut d'Astrophysique et G\'eophysique, Universit\'{e} de Li\`{e}ge, all\'{e}e du 6 Ao\^{u}t 17, B-4000 Li\`{e}ge, Belgium}

\altaffiltext{7}{Visidyne, Inc., 111 South Bedford St., Suite 103, Burlington, MA 01803, USA; devore@visidyne.com}

\altaffiltext{8}{Dipartimento di Fisica, Universit\'a di Torino, via P. Giuria 1, I-10125, Torino, Italy}

\altaffiltext{9}{Landessternwarte K\"onigstuhl, Zentrum f\"ur Astronomie der Universit\"at Heidelberg, K\"onigstuhl 12, D-69117 Heidelberg, Germany}

\altaffiltext{10}{Okayama Astrophysical Observatory, National Astronomical Observatory of Japan, Asakuchi, Okayama 719-0232, Japan}

\altaffiltext{11}{Institut de Ci\`encies de l'Espai (CSIC-IEEC), Campus
UAB, Facultat de Ci\`encies, Torre C5, parell, 2a pl., E-08193
Bellaterra, Spain} 

\altaffiltext{12}{Vanderbilt University, Nashville, TN 37235}
\altaffiltext{13}{Fisk University, Nashville, TN 37208}

\altaffiltext{14}{Stellar Astrophysics Centre, Department of Physics and Astronomy, Aarhus University, Ny Munkegade 120, DK-8000 Aarhus C, Denmark}

\altaffiltext{15}{Department of Geology \& Geophysics, University of Hawaii, 1680 East-West Road, Honolulu, HI 96822, USA and Visiting Astronomer at the Infrared Telescope Facility at the University of Hawaii.}

\altaffiltext{16}{Department of Earth and Planetary Sciences, Tokyo Institute of Technology, 2-12-1 Ookayama, Meguro-ku, Tokyo 152-8551, Japan}

\altaffiltext{17}{Harvard-Smithsonian Center for Astrophysics, 60 Garden St., Cambridge, MA 02138}

\altaffiltext{18}{Institute for Astronomy, University of Hawaii, 2680 Woodlawn Drive, Honolulu, HI 96822, USA}

\altaffiltext{19}{Harlan J. Smith Fellow}

\altaffiltext{20}{Department of Astronomy, The University of Texas at Austin, Austin, TX 78712, USA}

\altaffiltext{21}{Univ. Grenoble Alpes,IPAG, F-38000 Grenoble, France}

\altaffiltext{22}{CNRS, IPAG, F-38000 Grenoble, France}

\altaffiltext{23}{National Astronomical Observatory of Japan, 2-21-1 Osawa, Mitaka, Tokyo 181-8588, Japan}
\altaffiltext{24}{SOKENDAI (The Graduate University for Advanced Studies), 2-21-1 Osawa, Mitaka, Tokyo 181-8588, Japan}
\altaffiltext{25}{Astrobiology Center, National Institutes of Natural Sciences, 2-21-1 Osawa, Mitaka, Tokyo 181-8588, Japan}

 \slugcomment{Accepted for publication in the {\it Astrophysical Journal}, 2015 September 1}

\begin{abstract}

We present the discovery of a transiting exoplanet candidate in the K2 Field-1 with an orbital period of 9.1457 hr: K2-22b.  The highly variable transit depths, ranging from $\sim$0\% to 1.3\%, are suggestive of a planet that is disintegrating via the emission of dusty effluents.  We characterize the host star as an M-dwarf with $T_{\rm eff} \simeq 3800$ K. We have obtained ground-based transit measurements  with several 1-m class telescopes and with the GTC.  These observations (1) improve the transit ephemeris; (2) confirm the variable nature of the transit depths; (3) indicate variations in the transit shapes; and (4) demonstrate clearly that at least on one occasion the transit depths were significantly wavelength dependent.  The latter three effects tend to indicate extinction of starlight by dust rather than by any combination of solid bodies.  The K2 observations yield a folded light curve with lower time resolution but with substantially better statistical precision compared with the ground-based observations.  We detect a significant ``bump'' just after the transit egress, and a  less significant bump just prior to transit ingress.  We interpret these bumps in the context of a planet that is not only likely streaming a dust tail behind it, but also has a more prominent leading dust trail that precedes it.  This effect is modeled in terms of dust grains that can escape to beyond the planet's Hill sphere and effectively undergo `Roche lobe overflow,' even though the planet's surface is likely underfilling its Roche lobe by a factor of 2.

\end{abstract}


\section{Introduction}

The {\em Kepler} mission (Borucki et al. 2010) has revolutionized the field of exoplanets, with some 4000 planet candidates discovered to date (Mullally et al. 2015), of which at least 1000 have been confirmed (Lissauer et al. 2014, Rowe et al. 2014). With the original objective of discovering Earth-size planets in the habitable zone of their host stars, the telescope was bound to also improve our understanding of close-in rocky planets (Jackson et al. 2009; Schlaufman et al. 2010). Indeed, the first {\em Kepler} rocky planet, Kepler-10b, had an orbital period of only 20 hr (Batalha et al. 2011). The smallest planet with a well measured mass and radius, Kepler-78b, also has a very short orbital period of 8.5 hr (Sanchis-Ojeda et al.~2013), which was instrumental in measuring its small mass of 1.7 Earth masses (Howard et al.~2013, Pepe et al.~2013). In spite of the falloff in the numbers of {\em Kepler} exoplanet candidates at short periods, there are 106 well vetted candidates with orbital periods shorter than one day (hereafter ``USPs''; Sanchis-Ojeda et al.~2014), and most of them seem to be smaller than twice the size of Earth. 

Not included among the above lists are two special transiting exoplanets that are thought to be disintegrating via dusty effluents (Rappaport et al.~2012; Rappaport et al.~2014).  In both cases it is inferred that the planets are trailed by a dust tail whose dynamics are influenced by radiation pressure on the dust grains. This leads to transit profiles characterized by a pronounced depression in flux {\em after} the planet has moved off of the stellar disk (i.e., a post-transit depression).  In the case of KIC 12557548b (Rapapport et al.~2012; hereafter `KIC 1255b') the transit depths range from $\sim$1.2\% down to $\lesssim 0.1$\% in an highly erratic manner, while for KOI 2700b (KIC 8639908; Rappaport et al.~2014) the transit depths are observed to be slowly decreasing in depth over the course of the fours years of {\em Kepler} observations.  The fact that these `disintegrating' planets are relatively rare (2 of 4000 {\em Kepler} planets) is likely due to the conditions required for their existence and detection, namely high surface equilibrium temperatures and very low surface gravity, and a possibly short disintegration lifetime of only 10-100 Myr (see, e.g., Rappaport et al.~2012; Perez-Becker \& Chiang 2013).  

The main Kepler mission had an abrupt ending when two reactions wheels failed by March 2013. The reaction wheels are very important to maintain the telescope pointing in a given direction, and the telescope could no longer point toward the original {\em Kepler} field. The problem was partially bypassed by designing a new mission, called ``K2'', in which the telescope would point toward a different field of view along the ecliptic plane every three months (Howell et al. 2014); the spacecraft stability is improved by equalizing the Sun's radiation pressure forces on the solar panels. The unfortunate demise of the reaction wheels that put an end to the main mission, also opened the possibility for new discoveries of planets orbiting brighter stars since thousands of new bright stars are observed in each field. 

In the short lifespan of this new mission, there have been several papers describing techniques to produce light curves (Vanderburg \& Johnson 2014; Aigrain et al.~2015; Foreman-Mackey et al.~2015), and planet discoveries like a super-Earth transiting a bright host star (HIP 116454, Vanderburg et al.~2015), a triple planet system orbiting a bright M-dwarf (K2-3, Crossfield et al.~2015), and a pair of gas giants near a 3:2 mean motion resonance (EPIC 201505350, Armstrong et al. 2015b), with almost 20 confirmed K2 planets discovered to date (Montet et al.~2015). There have also been several catalogs of variable stars and eclipsing binaries (Armstrong et al.~2015a; LaCourse et al.~2015). This paper is the first in a series describing our discoveries using the K2 public data releases. The name of the project, ``ESPRINT'', stands for "Equipo de Seguimiento de Planetas Rocosos INterpretando sus Tr\'ansitos", which in English means "Follow-up team of rocky planets via the interpretation of their transits". 

In this work we focus on the surprising discovery of another one of these candidate disintegrating planets, this one in the K2 Field 1 which contains only 21,647 target stars (close to an order of magnitude fewer than in the prime {\em Kepler} field).  Even more impressive, this particular short-period exoplanet appears to have a dominant {\em leading} dust tail and possibly an additional trailing one, a phenomenon not seen before in astrophysics. The paper is organized as follows. In Section~\ref{sec:K2} we summarize the observations taken with the K2  mission and describe how this particular object was found.  In Section~\ref{sec:variability} we describe the variable transit depths, the timing analysis, and the unusual transit profile that cannot be explained by a solid body.  We present and discuss 15 transit measurements that were made in follow-up ground-based observations in Section~\ref{sec:transits}.  We analyze the properties of the host star based on a number of ground-based imaging and spectral observations in Section~\ref{sec:opthost}.  In Section \ref{sec:RVs} we set significant constraints on the radial velocity variations in the host star.  We discuss the wavelength dependence of the transits observed with the GTC in Section \ref{sec:GTC}.  We summarize why the host of the transits is the bright target star and not its much fainter companion in Section \ref{sec:brightstar}.  In Section~\ref{sec:dusttail} we interpret all the observations in terms of a model in which the planet is disintegrating, and discuss why the different characteristics and environment of K2-22b could lead to a dominant {\em leading} dust tail. Finally, in Section~\ref{sec:concl} we present a summary and conclusions and point toward new lines of research that could improve our understanding of how disintegrating planets form and evolve. 

\section{K2 data processing}
\label{sec:K2}

The target star, with EPIC number 201637175 (from now on named K2-22), was selected as one of the 21,647 stars in Field 1 to be observed in the long cadence mode of the K2 mission (Howell et al.~2014).  During the period from 2014 May 30 to 2014 August 20, a total of 3877 images of $15 \times 15$ pixels were recorded by the {\em Kepler} telescope, with a typical cadence of 29.42 minutes. The data were sent to NASA Ames, subsequently calibrated, including cosmic ray removal (Howell et al.~2014), and uploaded to the public K2 MAST archive in late 2014 December.  The data were then downloaded from the MAST archive and utilized for the analysis presented in this work. 

The discovery of K2-22b is part of the larger ESPRINT collaboration to detect and quickly characterize interesting planetary systems discovered using the K2 public data. In this section we highlight the way in which we produce light curves for all the observed stars, how this object was identified as part of the survey, and how we produced a better quality light curve for this particular object once the transits had been detected. 

\subsection{ESPRINT Photometric Pipeline}

Our photometric pipeline follows the steps of similar efforts published to date (Vanderburg \& Johnson 2014; Crossfield et al. 2015; Aigrain et al. 2015; Foreman-Mackey et al. 2015; Lund et al. 2015) that describe how to efficiently extract light curves from the calibrated pixel level data archived on MAST. The ingredients to generate the light curves, with our own choice described, are: 
\begin{itemize}
\item Aperture selection: Our apertures have irregular shapes, which are based on the amount of light that a certain pixel receives above the background level. We selected this type of aperture to capture as much light as possible while reducing the number of pixels used, which in turn reduces the noise induced by a large background correction. Based on experiments we carried out on the engineering data release,  selected pixels must be 30\% higher than the background level (estimated from an outlier corrected median of all the pixels in the image) in 50\% of the images in the case of a star brighter than {\em Kepler} magnitude $K_p =11.5$. For stars fainter than magnitude $K_p =14$, the selected pixels must be 4\% higher than the background. For stars of intermediate magnitudes, a linear interpolation of these two thresholds is used.  A simple algorithm groups contiguous pixels in different apertures , and the target star aperture is selected to be the one that contains the target star pixel position (obtained from the FITS headers). This type of aperture is similar to the ones used by Lund et al.~(2015), and quite different from the circular apertures used in many of the other pipelines. 

\item Thruster event removal: As highlighted in Vanderburg \& Johnson (2014), every 6 hours the telescope rolls to maintain the targets on the defined set of pixels that are downloaded, in what is known as a `thruster event'. We recognized these events by calculating the centroid motion of a particularly well behaved star (EPIC  201918073), and selected those moments where the position of the star jumps much more than usual (in the case of these stars, 0.1 pixels in the x direction). The images obtained during thruster events are removed from the analysis. 
\item Data slicing: We split our dataset into eleven different segments chosen to have a length of approximately 7 days, but also to contain an integer number of telescope roll cycles. The first segment and the one after a large gap (in the middle of the dataset) are not used in the global search since they are poorly behaved in some cases, with systematic effects induced by thermal changes that appear after reorienting the telescope.   
\item Systematics removal: We calculate the centroid positions and obtain a fourth-order polynomial that describes the movement of the star in the x and y coordinates. This polynomial fit is used to determine a new set of coordinates, in which the star moves only along one direction. The fluxes are then decorrelated first against time and then against this moving coordinate, using a fourth-order polynomial in each case. This process is repeated 3 times, and the results are very robust against problems caused by the presence of low-frequency astrophysical sources of noise (see Vanderburg \& Johnson 2014 for a more detailed description on how this process works).
\end{itemize}

\begin{figure}[h]
\begin{center}
\includegraphics[width=0.48 \textwidth]{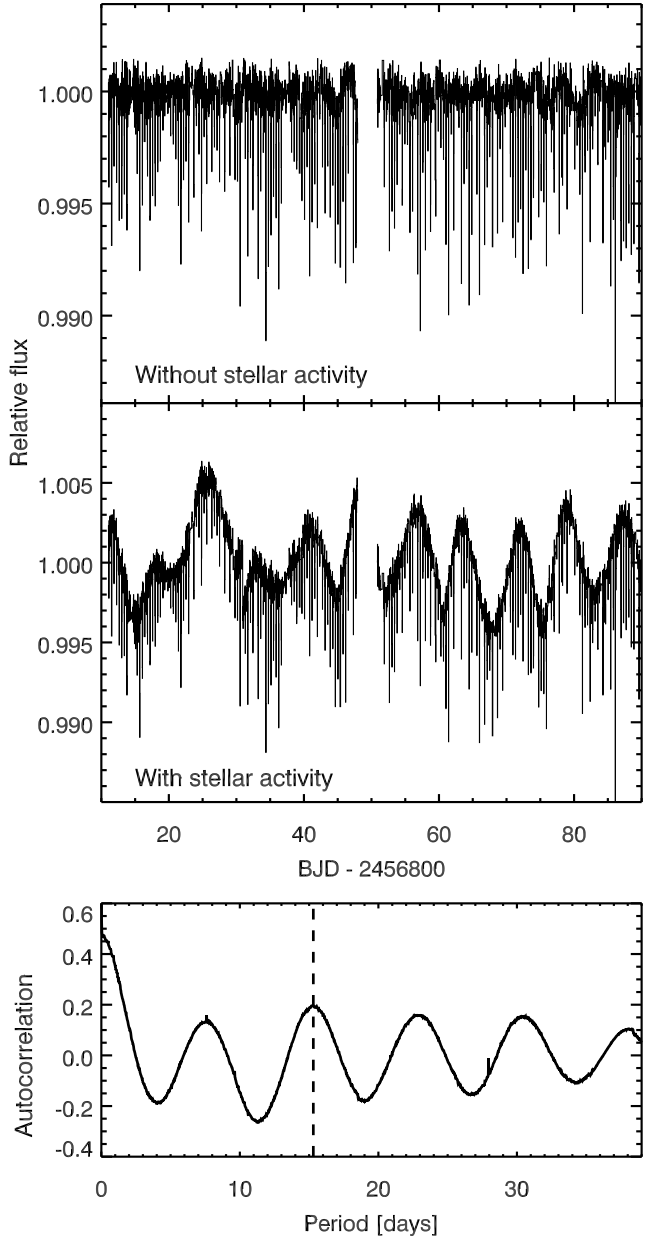} 
\caption{Lightcurve of K2-22 in time bins equal to the half-hour {\em Kepler} long-cadence sampling time.  Top panel: Light curve of the form used in our global search for USP planets. Middle panel: Light curve processed with a modified algorithm that better preserves stellar activity. Lower panel: Autocorrelation function, with a vertical line representing the inferred rotation period of the star.}
\label{fig:LC}
\end{center}
\end{figure}

This recipe was followed to generate the light curves of the 21,647 stars in Field-1. Among them, the light curve for EPIC 20163717 can be seen in the top panel of Fig.~\ref{fig:LC}. It is interesting to note that since our apertures depend on the amount of background light, which is increasing over the course of the observations, the number of pixels in the aperture decreases with time. This is a desired effect, since it tends to balance the increase of background light in the aperture by reducing the number of pixels, and therefore changes in the flux scatter are less severe. 

These light curves are generally analyzed using two different search algorithms: a more standard BLS routine (Kov\'{a}cs et al.~2002; Jenkins et al.~2010; Ofir 2014) to search for planets with orbital periods longer than 1 day, and a more specialized FFT pipeline used to detect planets with orbital periods shorter than 1-2 days (Sanchis-Ojeda et al. 2014). In this case, due to the short orbital period of the signal, we describe only the FFT search.

\subsection{Detection via the FFT technique}
\label{sec:detect}

Our target star K2-22 was detected as part of our search for ultra-short period planets in the K2 Field-1 dataset using the FFT technique. The routine automatically identifies those objects for which a main frequency and at least one harmonic can be distinguished above the level noise in the FFT power spectrum (see Sanchis-Ojeda et al. 2014 for details). A total of 2628 objects were identified in that way, but a large fraction of them were caused by improper corrections of the 6-hour roll of the {\em Kepler} telescope. These false detections are easy to remove since their main frequency is always related to the fundamental roll frequency of 4.08 rolls per day, although this simple removal clearly affects the completeness of our search. A total of 390 objects were selected for visual inspection, and among them K2-22 was selected as the most promising ultra-short orbital period planet candidate.

\begin{figure}
\begin{center}
\includegraphics[width=0.48 \textwidth]{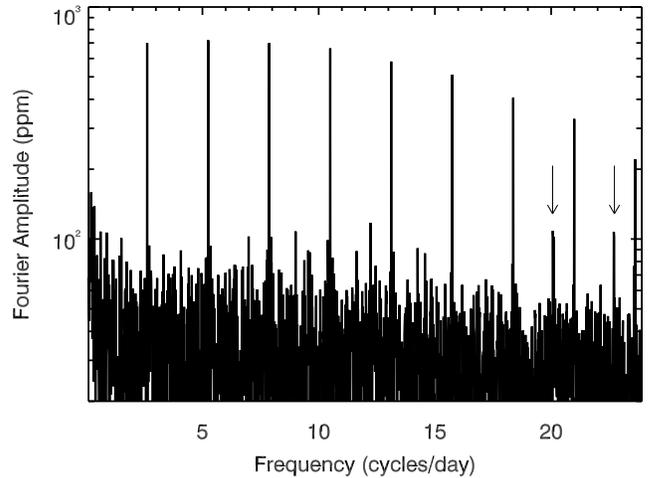}
\caption{Discovery Fourier transform of the flux data showing strong peaks at the 9.147-hour period and all 8 harmonics that are below the Nyquist limiting frequency. The arrows mark two additional harmonics aliased around the Nyquist limit.}
\label{fig:FT}
\end{center}
\end{figure}

The FFT by which this object was discovered is shown in Fig.~\ref{fig:FT}.  Note the prominent peak at 2.62 cycles/day which is the base frequency corresponding to the 9.1457-hour period, as well as the next 8 higher harmonics which lie below the Nyquist limit. The overall slowly decaying Fourier amplitudes with harmonic number is characteristic of short-period planet transits.
	
\subsection{Individualized aperture photometry}
\label{sec:aperture}

Our method for generating the light curves relies on a one-to-one relationship between the raw flux counts and the position of the star on the CCD chip. Any source of astrophysical variability could distort this relationship, and this is the case for both stellar activity induced signals and transits. A closer inspection of the raw light curve of K2-22 shows long-term trends that do not correlate with the centroid motion, and are likely due to the slow rotation of the host star. These trends are removed automatically as we fit a fourth-order polynomial in time to each of the 7-day segments, which effectively removes variability on scales longer than approximately 2 days (see upper panel of Fig.~\ref{fig:LC}).

\begin{figure}[h]
\begin{center}
\includegraphics[trim = {2cm 2cm 2cm 2cm}, clip, width=0.4 \textwidth]{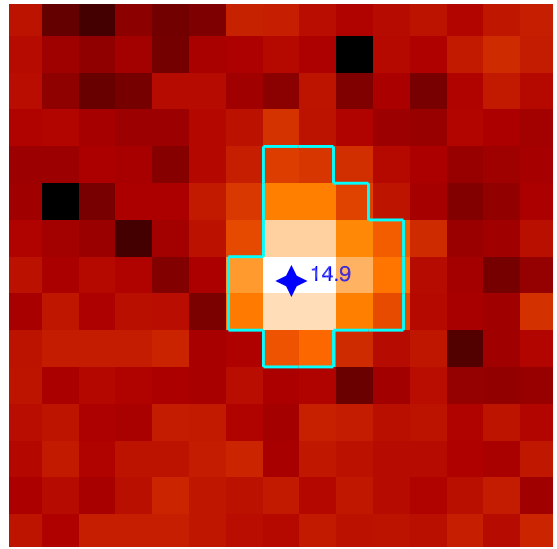}
\caption{K2 image of the object K2-22. A customized aperture is defined based on the amount of light of each pixel, and level of background light. The blue star represents the expected position of the target star, and the {\em Kepler} magnitude obtained from the EPIC catalog.}
\label{fig:K2image}
\end{center}
\end{figure}

In order to remove the effects of the transits, we first folded the original light curve given the period obtained from the FFT, after which we identified those orbital phases where no transit is expected.  We then ran our photometric pipeline again, using the same aperture (see Fig.~\ref{fig:K2image}) but only using the out-of-transit flux measurements to find the best fit polynomials to correct for both temporal and telescope motion variations. This process reduced the photometric scatter, and encouraged us to try different approaches to continue improving the light curve. We tried different combinations of polynomial orders and also different approaches to defining the apertures, but none of them improved the quality of the light curve (see top panel of Fig.~\ref{fig:LC}). After all the corrections, the final scatter per 30 minute cadence is 650 ppm, which is near the mean uncertainty obtained with our photometric pipeline for the typical K2 15$^{\rm th}$ magnitude star. 

We also tried to produce a light curve in which other astrophysical signals would be preserved (e.g., starspot rotation). During the process of detrending each of the 11 segments of data, we saved the coefficients of the fourth-order polynomial in time, and used them to reconstruct the signal again after removing the centroid motion artifact. Since the aperture is individually defined for each segment, we had to adjust the mean flux level of each segment to create a continuous light curve. This astrophysically more accurate light curve is also shown in Fig.~\ref{fig:LC}, and exhibits a clear signal of starspots with a rotation that could either be 7-8 days or twice this value. The shorter quasi-periodicity would typically arise when the star has two active longitudes separated by 180 degrees in longitude. We used an autocorrelation function to confirm this suspicion (see lower panel of Fig.~\ref{fig:LC}), and measured a rotation period of 15.3 days, following the techniques described in McQuillan et al. (2013).

\section{Transit parameters and depth variability}
\label{sec:variability}

In this section we describe the transits of K2-22b as observed by K2, with particular emphasis on the characteristics that deviate from the transits of a more typical planet.

\subsection{Individual transit times and depths}

\begin{figure}[t]
\begin{center}
\includegraphics[width=0.48 \textwidth]{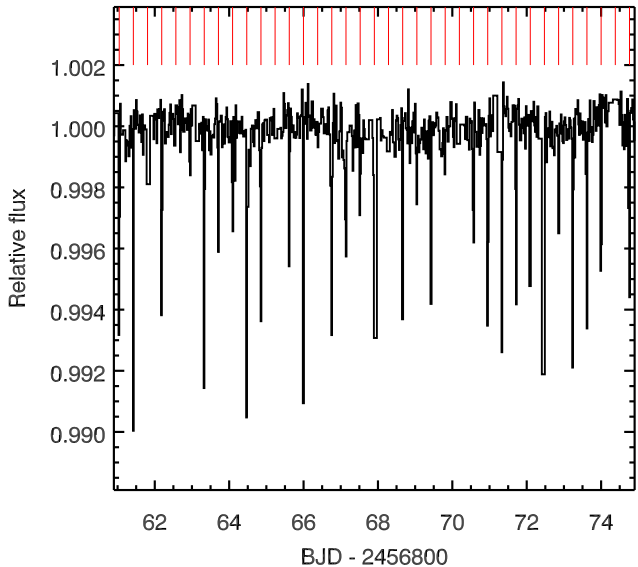} 
\caption{Zoomed version of Fig.~\ref{fig:LC} where the variability of the transit depths in much easier to see. Each vertical red line represents an expected transit time. }
\label{fig:zoom}
\end{center}
\end{figure}

The top panel of Fig.~\ref{fig:LC} shows the full data set from the K2 Field-1 observations covering an interval of $\sim$80 days.  It is apparent that there are sharp dips in intensity whose depths are highly and erratically variable.  In Fig.~\ref{fig:zoom} we can see a zoom in on two weeks of observations. The individual transits are now quite apparent, and the depth variations are dramatically evident.

 In order to analyze these variations, we first folded the light curve with the period obtained from the FFT, after removing any long-period signals.  We did this by fitting for local linear trends using a {\em total} of 5 hours of observations right before and after each transit.  We then fit the folded light curve with a simple idealized transit profile comprised of 3 straight-line segments: a flat bottom and sloping ingress and egress with the same slope magnitudes, hereafter referred to as the ``three-segment (symmetric) model''. This is very similar to the simpler `box model', but with non-zero ingress and egress times. We forced the fit to have a duration of at least 5 minutes on the lowest part of the transit, to make sure that the final fit did not look like a triangle.  This fit gave a mean transit depth of 0.5\%, and a total duration of 1 hour and 10 minutes. Given that the cadence of the observations is approximately 30 minutes, the real mean duration of the K2 transits must be close to 40 minutes, in agreement with what is expected of an USP planet orbiting an M-dwarf.  This is confirmed by our follow-up ground-based observations with their better temporal resolution (see Sect.~\ref{sec:transits}).

\begin{figure*}[ht]
\begin{center}
\includegraphics[width=0.95 \textwidth]{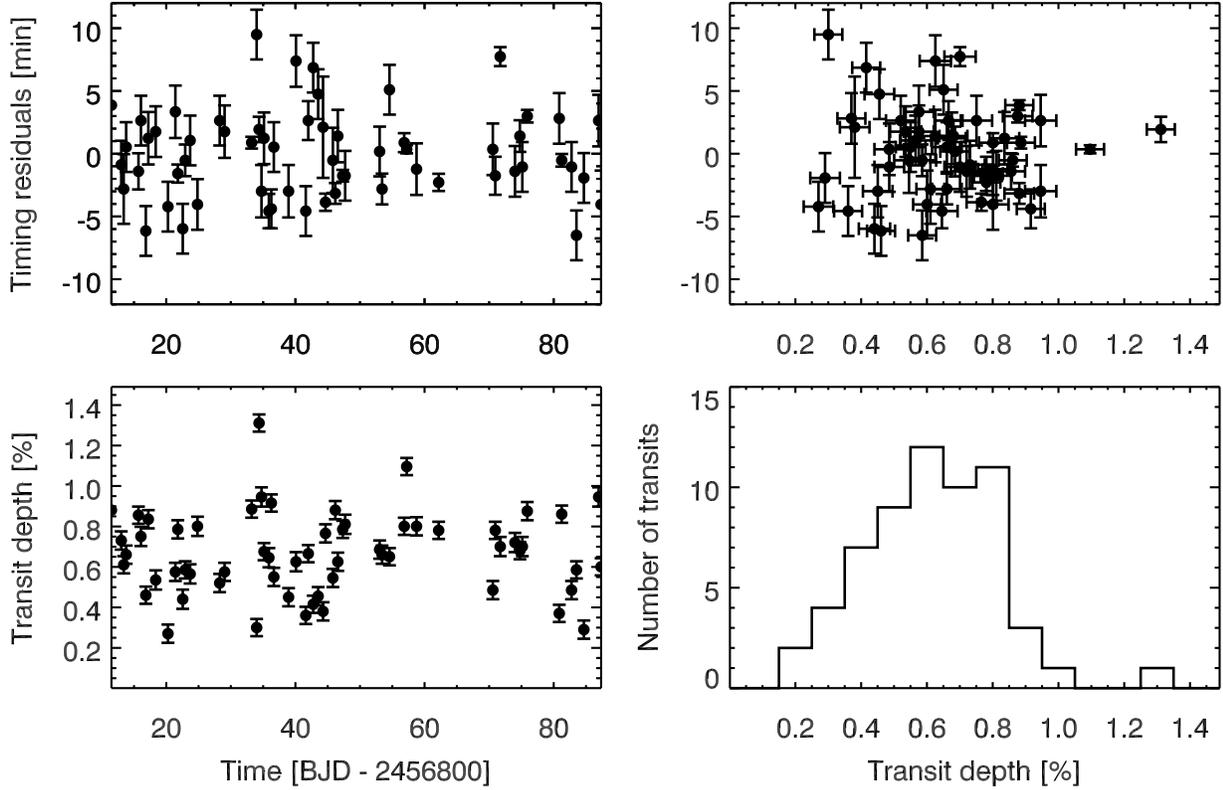}
\caption{In depth analysis of individual K2 transits. Upper left panel: Residuals of the K2 transit {\em times} with respect to a linear ephemeris, where the error bars reflect only the formal uncertainties. The best solution has a final $\chi^2$ of 480 for 60 transit times. Our simulations show that this excess of scatter comes from systematic uncertainties induced by the poor time resolution compared to the duration of the K2 transits. Lower left: The individual transit {\em depths} with time; these confirm the erratic variations in the transit depths. Upper right: Scatter plot showing that there is no significant correlation between the timing residuals and the transit depths. Lower right: the distribution of the 60 well-measured transit depths. This distribution is biased towards deeper transits, as transits shallower than $0.2\%$ were not analyzed due to the low S/N. }
\label{fig:k2trans}
\end{center}
\end{figure*}

With the mean transit profile and a good estimate for the orbital ephemeris, we fit each of the 190 individual K2 transits with the same three-segment model, allowing only the transit time to vary. In the process, we evaluated the robustness of the detection of each of the individual transits. In some cases, the depth is so low that the transit cannot be detected, whereas on other occasions the transit occurs during a thruster event, so there are no data. In rare cases, the transit consists of only a single flux point, which does not allow for a clear determination of the transit time or depth. After removing all those cases, we were left with a reduced sample of 60 well detected transits. In addition to the best fit transit time, we estimated the formal uncertainty by finding the interval of times where the standard $\chi^2$ function is within 1 of its minimum value, where a constant value of 650 ppm is used for the uncertainties in the flux. The transit times (including uncertainties) obtained in this manner are fit to a linear function, and the $O-C$ (`observed minus calculated') residuals of that fit are displayed in Fig.~\ref{fig:k2trans}. A clear excess of scatter (above the formal statistical uncertainties) is detected, with a best-fit standard $\chi^2$ of 480 for 60 transit times. Formal uncertainties, however, underestimate the true uncertainties when the model parameters are correlated, as could be the case here since we have removed a local linear trend for each transit which is known to be correlated with the transit time. We further examine this excess scatter below with simulations of the data train.

We now use the new orbital ephemeris to fix the time of each transit, and repeat the process but now fitting for the transit depths. These transit depths are significantly different from one transit to the next, and the depths range from a maximum of 1.3\% to 0.27\% which is close to the photometric detectability limit of our time series for individual flux points.  We can also see in Fig.~\ref{fig:k2trans} that the transit depths do not strongly correlate with the $O-C$ timing residuals. These erratic transit depth variations are quite reminiscent of those exhibited by KIC 1255b (Rappaport et al.~2012; Croll et al.~2014), and cover much the same range in depths.

We checked for periodicities in the measured transit depths and timing residuals using a Lomb-Scargle periodogram, but no significant peak was detected with a false alarm probability smaller than 1\%. We also attempted to constrain the true size of the planet by studying the shallowest K2 transits. We revisited the discarded transits, and selected the 6 shallowest cases in which the transit observations were complete (no thruster events). The mean depth of these transits is  $0.14 \pm 0.03 \%$, which, given the radius of the star (see section~\ref{sec:spectraana}), translates into an upper bound of $2.5 \pm 0.4$ $R_\oplus$ on the planet radius.

Even though the transit depth variations described above are later confirmed by ground-based observations, there is a potential concern that the variability could have been caused by the relatively comparable transit and sampling timescales. In this same regard, it is also possible that some or all of the excess variations in the $O-C$ scatter (see Fig.~\ref{fig:k2trans}) over those expected from statistical fluctuations are due to the relatively short transit duration compared with the LC integration time.  We have therefore carried out extensive numerical simulations of these effects. 

The simulations of the depth and transit-time variations are based on a model that has a simple box transit profile which is integrated over the 30 minute LC time, and includes a sinusoid (and its first harmonic) to represent the rotating starspot activity. The amplitude of the sinusoids and the rotation period are fixed at 1\% and 15 days, respectively. The duration of the model transit is fixed at 50 min (see Sect.~\ref{sec:transits}). The model is evaluated at the same times as the K2 observations, and only the same 60 transit windows are analyzed to ensure that the simulations represent a similar dataset to the one used in this paper. There are a total of four different numerical experiments, each one repeated many times, in which we either include starspots or not, and we either have a constant transit depth or depths drawn from a Gaussian distribution of a given variance, but always with a mean of 0.5\%. In all cases a fixed orbital ephemeris is used.  White noise of 700 ppm per LC sample is added.  The simulated datasets are then processed with the same pipeline used as was used in this work for the K2 data. 

The main conclusions from these simulations are: (i) The formal uncertainties in the `measured' transit depths ($\sim$0.05\%) are indeed underestimates of the actual uncertainties. The simulations using a constant depth have recovered depths with a scatter of 0.1\%. This is likely the result of a combination of systematic effects induced by the short duration of the transits and the white noise terms.  The simulations show that an actual scatter in the depths greater than 0.15\% is easily recoverable. The scatter of our K2 depths is 0.2\%, so we conclude that it is real. The ground based observations confirm this.  (ii) The formal uncertainties also underestimate the true uncertainties of the transit times (defined as the scatter of measured times after removing a best linear trend). This effect is worse in the presence of starspots, but it does not depend on the transit depth variations. The effect can be large enough to explain {\em all} the scatter observed in the K2 timing analysis and therefore this scatter is {\em not} significant. We have added a systematic uncertainty in quadrature to the timings of 3 minutes, chosen to provide a best-fit reduced $\chi^2$ of 1.  (iii) Our treatment of stellar spots does not induce detectable depth variations (see Kawahara et al.~2013; Croll et al.~2015), as expected, mostly because we do not have the precision to detect them.

\subsection{Deviation from a standard transit profile}
\label{sec:diverge}

\begin{figure}[h]
\begin{center}
\includegraphics[width=0.48 \textwidth]{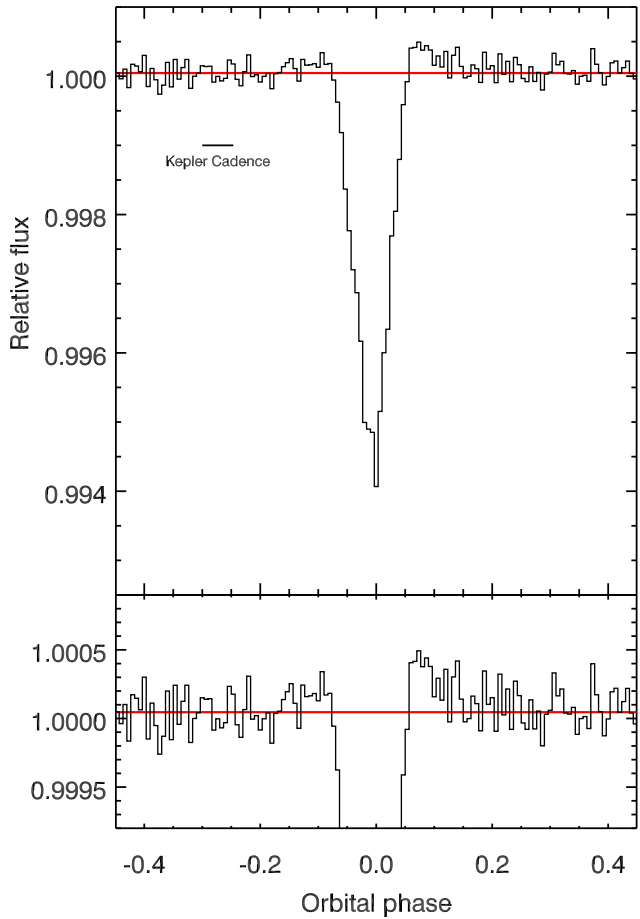} 
\caption{Folded light curve of K2-22b for an orbital period of 9.14570 hours.
The folded data have been averaged into 3.7-minute bins.  The empirical out of transit rms variations in the flux are 195 ppm. As explained in the text there is a positive ``bump'' in flux just after the transit egress and a smaller, but still marginally significant, bump just prior to ingress. The red horizontal lines represent the mean out-of-transit normalized flux. The black horizontal bar indicates the LC time of 29.4 minutes, and gives an indication of the inherent temporal resolution of the light curve}
\label{fig:fold}
\end{center}
\end{figure}

A fold of the activity corrected data (see section~\ref{sec:aperture}) about the period we determined of 9.145704 hours is presented in Fig.~\ref{fig:fold}.  The overall crudely triangular shaped transit profile is the result of a convolution of the intrinsic shape and the LC sampling time (see also Sects.~\ref{sec:transits} and \ref{sec:GTC}). The depth of the folded profile is 0.6\%, and of course, this represents an average of the highly variable depths.  The red curve is the mean out-of-transit normalized flux (averaged for orbital phases between $-0.5$ to $- 0.25$, and from 0.25 to 0.5).  Note the clear positive ``bump'' in flux just after the transit egress and the smaller, but still marginally significant, bump just prior to ingress. These are significant at the 6-$\sigma$ and 2-$\sigma$ confidence limits, respectively.

These ``bumps'', which are not normal features of exoplanet transits, will be important for understanding the basic nature of the transits (see Sect.~\ref{sec:dusttail}).  We call these features the ``pre-ingress bump'' and the ``post-egress bump''.  Even in the cases of KIC 1255b and KOI 2700b, the other two exoplanets which appear to have dusty tails, the main distinguishing feature of their transit profiles is a post-transit {\em depression} (Rappaport et al.~2012; Rappaport et al.~2014) which is attributed to a {\em trailing} comet-like dust tail.  In addition, KIC 1255b exhibits a {\em pre}-ingress ``bump'' which has been attributed to forward scattering in the dust tail near the head of the dust cloud (see, e.g., Rappaport et al.~2012; Brogi et al.~2012; Budaj 2013; van Werkhoven et al.~2014).  By contrast, the transit profile of K2-22b has no post-egress depression, and the most prominent bump comes {\em after} the egress, rather than before.  These features will be crucial to the interpretation of the dust ``tail'' in this system. 

Finally,  even though we believe the transit is due to a dust tail, as a baseline reference model we attempted to fit a standard transit profile of a solid planet over a limb-darkened star (Mandel \& Agol~2002) to our folded light curve. The high distortion of the transit light curve due to the 30 minute sampling precludes obtaining precise transit parameters, but we were able to constrain the scaled semi-major axis from the transit itself  to be $d/R_* = 4.2^{+0.15}_{-0.5}$, with the uncertainties estimated from an MCMC analysis.  In fact, from the stellar properties obetained in section~\ref{sec:spectraana}, and Kepler's 3rd law, we can make a better direct estimate of $d/R_* = 3.3 \pm 0.2$, which is compatible with the transit fit. We ran a final transit model with a Gaussian prior on $d/R_*$ based on the inferred stellar density, with mean value of 3.3 and a standard deviation of 0.2.  This, in turn, allowed us to estimate a mean K2 total transit duration (first to fourth contact) of $46 \pm 1$ minutes, an impact parameter of $b = 0.68 \pm 0.06$ (see Table~\ref{tbl:planet}) and a mean depth of $(R^\prime_p/R_*)^2 = 0.55$\%, where $R^\prime_p$ can be understood as the mean effective radius of the  dust grains. 

\subsection{Verification With Independently Processed K2 Data}

We have also used the Vanderburg \& Johnson (2014) flux time series data set for K2-22 to check against the results of our own pipeline.  With the application of a simple high-pass filter, this independently processed light curve looks nearly identical to the one shown in the upper panel of Fig.~\ref{fig:LC}.  A fold of the data about the period we have determined yields a transit profile that is essentially the same as that shown in Fig.~\ref{fig:fold}, including the appearance of a convincing post-egress bump, and a somewhat less significant pre-eclipse bump.  To the extent that our pipeline and that of Vanderburg \& Johnson (2014) are independent, this is a satisfying test that our processing has introduced no artifacts into the lightcurve.

\section{Ground-Based Transit observations}
\label{sec:transits}

A total of 12 transits were obtained from the ground using 1-m class telescopes, and the observations are summarized in Table~\ref{tbl:log}. The first observations were taken with the 1.2-meter telescope at the Fred Lawrence Whipple Observatory (FLWO) on Mt. Hopkins (AZ) using the KeplerCam instrument, which has a single 4K $\times$ 4K Fairchild 486 CCD with a $23'.1 \times 23'.1$ field of view.  We successfully observed three different transits with photometry in the Sloan i' filter. 

We also conducted a follow-up transit observation with the Okayama 1.88-m telescope using the NIR camera ISLE (Yanagisawa et al.~2006), and adopting a similar observing scheme to that described in Fukui et al.~2014, we studied the transit of the target in $J$-band. We only slightly defocused the stellar images, but due to the faintness of the target and reference star, the raw counts were well within the detector's linear range.  The typical FWHM of the target star's PSF was $\sim 14$ pixels, which corresponds to $\sim 3.5^{\prime\prime}$. 

We obtained two additional transits with the IAC-80 (80-cm) telescope at 
the Observatorio de Izaña, in the Canary Islands. We used 
the wide field CAMELOT camera, with a FOV of $10.4' \times 10.4'$, with 
observations taken in the i-band. No defocussing was applied. Weather 
conditions were clear and stable through the two nights.

A total of six transits were observed with the 0.6-m TRAPPIST robotic telescope (TRAnsiting Planets and PlanetesImals Small Telescope), located at ESO La Silla Observatory (Chile). TRAPPIST is equipped with a thermoelectrically-cooled 2K $\times$ 2K CCD, which has a pixel scale of 0.65$''$ that translates into a 22$' \times 22'$ field of view. For details of TRAPPIST, see Gillon et al. (2011) and Jehin et al. (2011). The observations were obtained through a blue-blocking filter\footnote{http://www.astrodon.com/products/filters/exoplanet/} that has a transmittance of $> 90\%$ from 500 nm to beyond 1000 nm, without any additional defocus due to the faintness of the target star. The procedures for the observation and data reduction are similar to those described by Gillon et al. (2013) and we refer to this paper for further details.

Finally, additional observations were taken in queue mode with OSIRIS@GTC on three different nights (see Table~\ref{tbl:log}), covering three complete transits of  K2-22b. A spectroscopic time series was taken in staring mode, starting $\approx 1$ h before the ingress, and finishing $\approx 1$ h after the egress. The observing logs are summarized in Table~\ref{tbl:log}. Even though the GTC data were taken in low-resolution spectral mode, for the purpose of transit timing, we used the data integrated over wavelength (i.e., an effective white light transit).  The spectral dependence of the transits is discussed in detail in Sect.~\ref{sec:GTC}.

In all, 15 transits were measured from the ground.  A log of these 15 observations is given in Table \ref{tbl:log}.

\begin{deluxetable*}{lccllccclccc}

\tablecaption{Ground based transit observations of K2-22b\label{tbl:log}}

\tablehead{
\colhead{Date} &
\colhead{Start} &
\colhead{End} &
\colhead{Epoch} &
\colhead{Telescope} &
\colhead{Instrument} &
\colhead{Number of} &
\colhead{Median time}&
\colhead{Airmass\tablenotemark{a}} & 
\colhead{$\sigma$\tablenotemark{b}} &
\colhead{$\beta$\tablenotemark{b}} \\
\colhead{[UT]} &
\colhead{} &
\colhead{} &
\colhead{} &
\colhead{} &
\colhead{or filter} &
\colhead{data points} &
\colhead{between points [min]}  & 
\colhead{} &
\colhead{ [ppm]} &
\colhead{} 
}

\startdata
2015 Jan 15 & 07:30 & 12:45 & 595 & 1.2 m FLWO & Sloan i' & 136 & 2.3 & $1.89 \rightarrow 1.15 \rightarrow 1.25$ & 2800  & 1 \\
2015 Jan 17 & 14:40 & 18:23 &  601  & Okayama & J-band & 102 & 2.1  & $2.32 \rightarrow 1.18 \rightarrow 1.69$ &  3000 & 1 \\
2015 Jan 23 & 07:13 & 12:06 &  616  & 1.2 m FLWO & Sloan i' & 126 & 2.3 & $1.78 \rightarrow 1.15 \rightarrow 1.24$ &  3500 & 1 \\
2015 Jan 27 & 02:33 & 05:55 &  626  & IAC-80 & Sloan i & 117 & 1.7  & $1.22 \rightarrow 1.13 \rightarrow 1.28$ & 4070  & 1.24 \\
2015 Jan 29  & 03:15  & 06:45  & 634 & GTC & R1000R &  44 & 4.6 &  $1.12 \rightarrow 1.11 \rightarrow 1.53$ & 570  & 1.13 \\
2015 Feb 02 & 03:24 & 09:35 &  642  & TRAPPIST & Blue blocking & 314 & 1.1  & $2.34 \rightarrow 1.18 \rightarrow 1.42$ & 5500  & 1.15  \\
2015 Feb 04 & 02:16 & 07:06 &  647  & IAC-80 & Sloan i & 260 & 1.1  &  $ 1.18 \rightarrow 1.11 \rightarrow 1.79$&  4300  & 1.00  \\
2015 Feb 12 & 02:32 & 06:28 &  668  & TRAPPIST &Blue blocking & 207 & 1.1  & $2.60 \rightarrow 1.18 \rightarrow 1.18$ & 4070  & 1.52 \\
2015 Feb 14  & 00:10  & 03:30 & 673 & GTC & R1000R & 40 & 5.4 &  $1.45 \rightarrow 1.11 \rightarrow 1.12$ &  920 & 1.46 \\
2015 Feb 15 &  03:24  & 06:42 & 676 & GTC & R1000R &  38 & 5.4 &  $1.12 \rightarrow 1.12 \rightarrow 2.07$ & 690  & 1.26\\
2015 Feb 18 & 05:22 & 12:30 &  684 & 1.2 m FLWO & Sloan i' & 188 & 2.3 & $1.88 \rightarrow 1.15 \rightarrow 1.90$ & 2270  & 1.50  \\
2015 Feb 23 & 03:27 & 07:39 &  697  & TRAPPIST & Blue blocking& 213 & 1.1  & $1.50 \rightarrow 1.18 \rightarrow 1.30$ & 2600  &  1.38\\
2015 Feb 25 & 02:27 & 05:24 &  702  & TRAPPIST & Blue blocking & 137 & 1.1  & $1.71 \rightarrow 1.18 \rightarrow 1.18$ & 3230  & 1 \\
2015 Feb 26 & 04:38 & 08:48 &  705  & TRAPPIST &Blue blocking & 202 & 1.1  & $1.23 \rightarrow 1.18 \rightarrow 1.69$ & 2760  & 1 \\
2015 Mar 21 & 02:00 & 05:37 &  765 & TRAPPIST &Blue blocking& 182 & 1.1 & $ 1.43 \rightarrow 1.18 \rightarrow 1.26$  & 2800  &  1
\enddata

\tablenotetext{a}{The airmass range is shown as $z_0 \rightarrow z_{\rm min} \rightarrow z_{\rm fin}$, where $z_0$ and $z_{\rm fin}$ represent the airmass at the beginning and at the end of the night, respectively, and $z_{\rm min}$ represents the minimum airmass. }
\tablenotetext{b}{$\sigma$ refers to the flux scatter respect to the best fit model, whereas $\beta$ represents the level of correlated noise (see section~\ref{sec:transits}).}

\end{deluxetable*}

Each of the above instruments has a different procedure for reducing the light curves, and we refer to the corresponding literature for a more detailed explanation (Okayama: Fukui et al.~2011; TRAPPIST: Gillon et al.~2013; FLWO: Holman et al.~2006; IAC80: L{\'a}zaro et al.~2015; GTC: see Sec.~\ref{sec:GTC}). All transit light curves were obtained by comparing the fluxes of the host star to a reference light curve made by combining up to several comparison stars. They were all adjusted for differential airmass corrections and a second-order polynomial was fitted to the out-of-transit part of the light curves to remove long-term trends. The times of observation for all light curves were transformed into BJD (TDB format, see Eastman et al.~2010) to compare them with the K2 observations.

\begin{figure*}[h]
\begin{center}
\includegraphics[width=0.95 \textwidth]{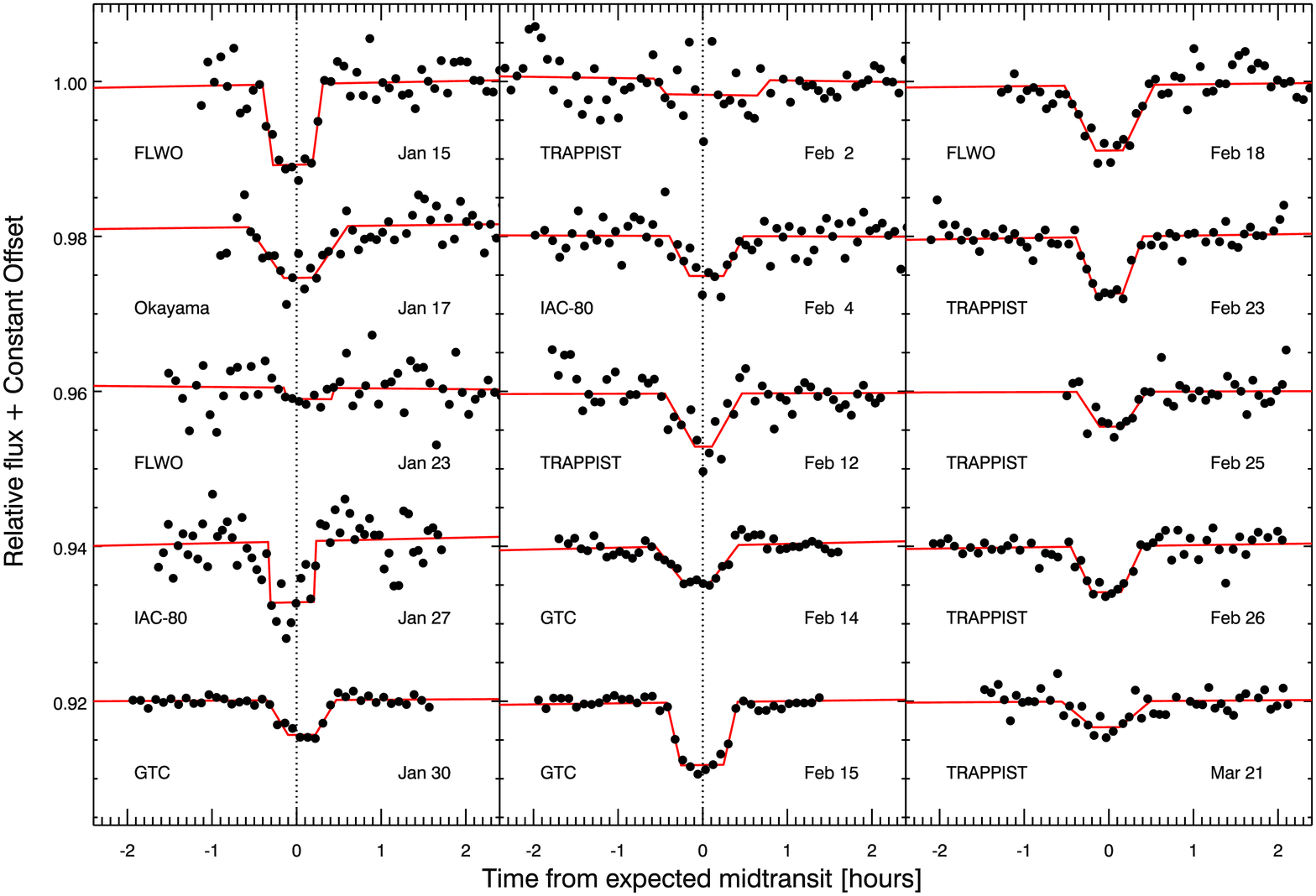}
\caption{Fifteen transits of K2-22b observed from the ground, plotted to the same vertical scale.  The transit models are each vertically offset by 0.02 for clarity in presentation. When necessary, the observations are binned to have a cadence close to 5 minutes. In spite of the weaker statistics for some of these, it is apparent that the transit depths vary considerably.}
\label{fig:groundtrans}
\end{center}
\end{figure*}

The 15 transit profiles measured from the ground are presented in Fig.~\ref{fig:groundtrans}.  The profiles were fit with the same simple three-segment transit model used in the K2 data analysis. The parameters were transit depth, time of transit, transit duration, and ingress time. We also added two parameters to fit for any fiducial linear trends with time. In several cases a transit fit barely represents an improvement over a straight-line fit, whereas in some cases a deep transit is detected, confirming the depth variations (see Fig.~\ref{fig:groundtrans}). We first scaled the uncertainties in the flux measurements to be equal to the standard deviation of the flux residuals with respect to the best fit model. Correlated noise was taken into account using the time-averaging method, in which the ratio of the standard deviation of the time-averaged residuals and the standard deviation expected assuming white noise is calculated over a range of timescales (Pont et al. 2006; Winn et al. 2008). In our case, we obtained the final values of this $\beta$ parameter as the mean of the ratios with timescales from 10 to 30 minutes (see Table~\ref{tbl:log}), and multiplied the initial uncertainty by $\beta$ when estimating the standard $\chi^2$ function. In those 9 cases where the transit fit improves the minimum $\chi^2$ by at least 35, an MCMC routine was used to estimate the uncertainties in the transit parameters. The 9 new transit times obtained from the ground based observations are summarized in Table~\ref{tbl:times}. We combined the K2 transit times with the new 9 transit times to obtain the final orbital ephemeris. The new transit times also show a high level of scatter ($\chi^2 = 30$ for 9 O-C determinations).

\begin{figure*}[ht]
\begin{center}
\includegraphics[width=0.9 \textwidth]{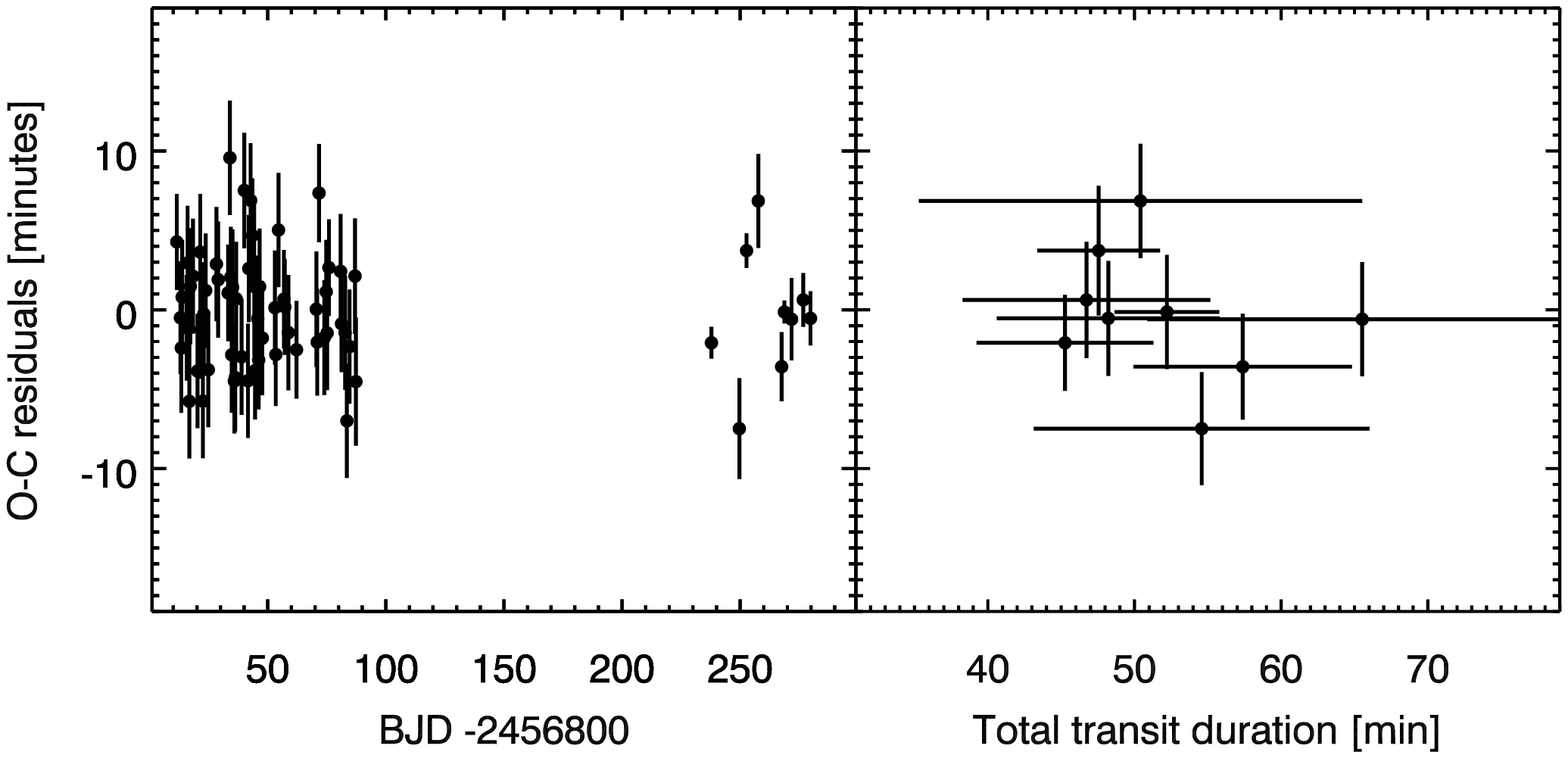}
\caption{Left panel: Timing residuals after removing the best fit linear orbital ephemeris. The uncertainties on the K2 timings have been increased to take into account systematic effects. Right panel: The uncertainties in the transit {\em durations} are too high to allow for a detection of a possible correlation between transit duration and transit time.}
\label{fig:timings}
\end{center}
\end{figure*}

The O-C values for all the well measured K2 and ground-based transits are summarized in Fig.~\ref{fig:timings} (left panel).  We used these data to determine a best fit orbital period and its uncertainty (after multiplying by the square root of the reduced $\chi^2$).  We repeated the process of fitting the O-C points, but this time with a quadratic function, to set upper bounds on the derivative of the orbital period.  The period and period derivative results are summarized in Table~\ref{tbl:planet}.

\begin{deluxetable}{ccc}

\tablecaption{Ground based transit times\label{tbl:times}}
\tablewidth{0pt}

\tablehead{
\colhead{Epoch} &
\colhead{Time of transit (BJD)}  & 
\colhead{Uncertainty (days)}
}

\startdata
  595    &    2457037.8608  &  0.0007 \\ 
  626    &    2457049.6705   & 0.0022 \\
  634     &   2457052.7269  &  0.0008 \\
  647    &    2457057.6831  &  0.0021 \\
  673     &   2457067.5839  &  0.0015 \\
  676     &   2457068.7295  &  0.0005 \\ 
  684     &   2457071.7778  &  0.0018 \\
  697     &   2457076.7327  &  0.0012 \\
  705     &   2457079.7805  &  0.0012 
\enddata
\tablenotetext{}{See section~\ref{sec:transits} for details.}
\tablenotetext{}{}

\end{deluxetable}

The extra source of scatter in the ground-based transit timings could be caused by changes in the shape of the transit light curve (see Croll et al. 2015). In principle these timing variations should be accompanied by transit duration or shape variations, but our light curves are not precise enough to allow the detection of such correlated variations (see right panel of Fig.~\ref{fig:timings}). We cannot discard the possibility that part of the scatter is due to the use of a symmetric transit profile to compute transit times, when indeed some of the transits appear slightly asymmetric. From this set of 9 high-quality ground-based transit measurements we obtained a weighted {\em total} transit duration average of $50 \pm 2$ min, slightly longer than the duration obtained from the K2 photometry. This difference of $4 \pm 2$ minutes, can be explained by the non-zero cadence of the ground-based observations, which has not been taken into account in these fits. A final value of $48 \pm 3$ minutes is quoted for the total transit duration in Table~\ref{tbl:planet}.

\section{Properties of the Host Star}
\label{sec:opthost}

\subsection{Imaging and color information}
\label{sec:image}

An early inspection of the SDSS images of the target star, K2-22, showed that it is quite cool and likely an M star.  The image also indicated the presence of a faint companion at a distance $\sim$2$''$ in the South-West direction. 

\begin{figure}[t]
\begin{center}
\includegraphics[trim = {0cm 0cm 0cm 0cm}, clip, width=0.4 \textwidth]{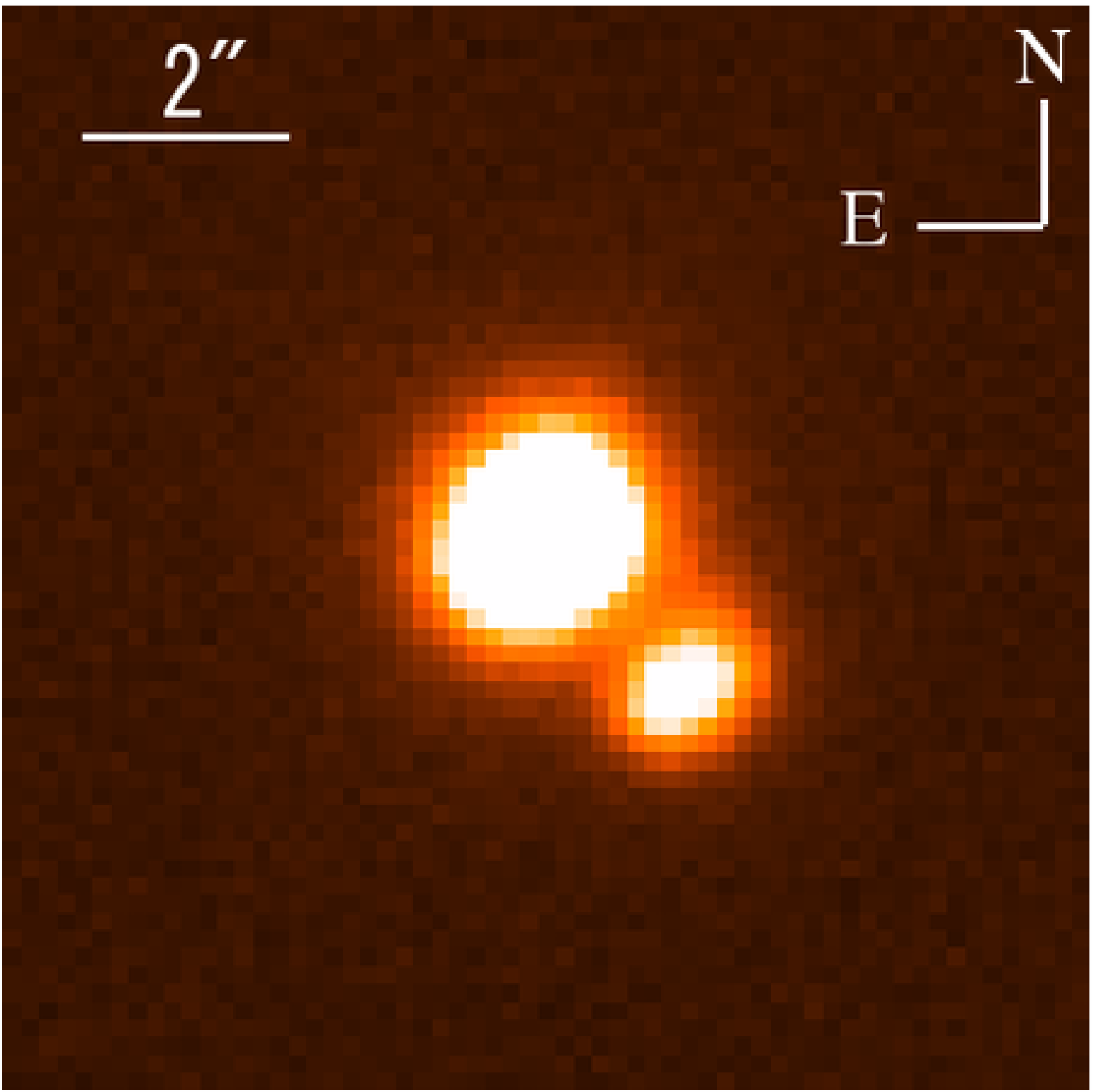}
\caption{Subaru/HSC+z image of K2-22 shows a secondary companion star $\sim 2^{\prime\prime}$ away, with a position angle of 134 degrees west from north. In this band, the flux ratio between the faint and bright star is 12\%.}
\label{fig:subaru}
\end{center}
\end{figure}

We observed the target star and its companion with Hyper-Suprime Cam (HSC, Miyazaki et al.~2012) on the Subaru 8.2-m telescope on 25 January, 2015 (UT). The sky condition on that night was clear and photometric. The HSC is equipped with 116 fully-depleted-type $2048 \times 4096$ CCDs with a pixel scale of 0.17$''$.  We took images of 3~s and 60~s exposures through g-, i-, and z-band filters (see Fig.~\ref{fig:subaru}). The host star  is saturated in i- and z-band with 60-s exposures, while the companion star is not clearly seen in the g-band image with 3-s exposure. We thus use the 60-s exposure images for g-band and 3-s exposure images for i- and z-band for the analysis presented here. 

\begin{deluxetable}{lcc}

\tablecaption{Properties of the host star and companion\label{tbl:stars}}
\tablewidth{0pt}

\tablehead{
\colhead{Parameter (units)} &
\colhead{Host star} &
\colhead{Companion} 
}

\startdata
RA (J2000) & 11:17:55.856 & 11:17:55.763 \\
DEC (J2000) & +02:37:06.79  & +02:37:05.48 \\
u-mag (SDSS) &  \,\,\,  $19.07 \pm 0.05$ \,\,\,  & \,\,\,   $21.68 \pm 0.17$ \,\,\,  \\
g-mag (SDSS) &  $16.44 \pm 0.05$ &  $19.61 \pm 0.05$ \\
r-mag (SDSS) &  $15.01 \pm 0.05$ &  $18.79 \pm 0.05$ \\
i-mag (SDSS) &  $14.38 \pm 0.05$ &  $17.34 \pm 0.05$ \\
z-mag (SDSS) &  $14.05 \pm 0.05$ &  $16.34 \pm 0.05$ \\
J (2MASS) & $ 12.74 \pm 0.05 $  & $14.87 \pm 0.05$ \\
H (2MASS) &  $12.09  \pm 0.05 $ 	&  $14.27  \pm 0.05 $  \\
$K_S$ (2MASS) & $11.91   \pm 0.05 $  & $  13.93  \pm 0.05 $  \\
$T_{\rm eff}$ (K) & $3830 \pm 100$ & $3290 \pm 120$ \\
$\log\,g$ & $4.65\pm0.12$ & ... \\
$ [Fe/H]$& $0.03\pm0.08$ &  $0.06\pm0.20$ \\
$M_* \, (R_\odot)$ & $0.60 \pm 0.07$ & $0.27 \pm 0.05$ \\
$R_* \, (M_\odot)$ & $0.57 \pm 0.06$ & $0.30 \pm 0.08$ \\
$L_* \, (L_\odot) $ & $0.063^{+0.008}_{-0.007}$ & $0.010^{+0.007}_{-0.005}$ \\
Spec~Type & M0V $\pm 1$ & M4V $\pm 1$ \\
Distance (pc) & $225 \pm 50$ & $225 \pm 50$ 
\enddata
\tablenotetext{}{The magnitudes and colors are taken from a combination of the SDSS and 2-MASS photometry.  The stellar parameters are inferred from the combined analyses of the Keck-HIRES, IRTF-SpeX, and UH88-SNIFS spectra. See section~\ref{sec:opthost} for details.}
\tablenotetext{}{}

\end{deluxetable}

The presence of a stellar companion can be problematic, particularly when the distance between the stars is smaller than the pixel size of {\em Kepler}. We measure the flux ratios of the companion to the host star in the HSC g-, i-, and z-band images (see Fig.~\ref{fig:subaru})  by using the DoPHOT program (Schechter et al.~1993), which performs PSF-fitting photometry while de-blending stars. The background star is situated at a distance of 1.91$''$, with a position angle of 134 degrees west from north (see Table~\ref{tbl:stars} for the coordinates). We obtain flux ratios of $4.0 \pm 0.5 \%$, $8.6 \pm 0.5 \%$, and $11.7 \pm 0.5 \%$  for the g, i, and z bands, respectively, where the uncertainties have been increased to take into account systematic sources of noise. 

The contribution of the background star points toward a cooler and fainter companion. The bright star has been characterized as part of the K2-TESS catalog, in which effective temperatures are obtained from the colors of the stars (Stassun et al.~2014). The temperature reported there for the brighter target star is $\sim 3700$ K, which means that both stars are likely to be dwarfs stars (see Sect.~{\ref{sec:spectraana}). The colors used in that analysis could have been blended due to the proximity of the companion.  We also obtained the $ugriz$ SDSS magnitudes and the J 2MASS magnitude for both stars, confirming that the colors used in Stassun et al.~(2014) were correct. We re-derived the flux ratios in g, i and z band (5.4\%, 6.5\% and 12\% respectively) confirming the measurements obtained using the Subaru images. The minimum uncertainties used in the fits are 0.05 magnitudes to take into account systematic effects. Adopting the full line-of-sight extinction of $A_V = 0.17$ based on the expected distance to the objects, and a $\log \, g$ of 5 due to their red colors and dwarf nature (see Sect.~{\ref{sec:spectraana}), we obtained temperatures of $3800 \pm 150$ K for the bright star and $3150 \pm 200$ K for its fainter companion.

The magnitudes and colors of the two stars based on a combination of the SDSS, Subaru, and 2-MASS photometry are summarized in Table \ref{tbl:stars}.

\begin{deluxetable}{lc}

\centering

\tablecaption{Properties of K2-22b \label{tbl:planet}}
\tablewidth{0pt}

\tablehead{
\colhead{Parameter} &
\colhead{Value} 
}

\startdata
Orbital Period\tablenotemark{a} (days) & $0.381078 \pm 0.000001$ \\
Orbital Period (hr) &  $9.145872 \pm  0.000024$  \\
Transit center\tablenotemark{a} (BJD) & $2456811.1208 \pm 0.0006$ \\
$\dot P_{\rm orb}/P_{\rm orb}$ (yr$^{-1}$)\tablenotemark{a} & $\lesssim 3.5 \times 10^{-7}$  \\
Total transit duration\tablenotemark{a} (min) &  $48 \pm 3$   \\
$d/R_{\rm *}$\tablenotemark{b} &  $3.3 \pm 0.2$   \\
Impact parameter, $b$\tablenotemark{c} & $0.68 \pm 0.06$ \\
$d \, (AU)$ & $ 0.0088 \pm 0.0008 $ \\
$R_p \,(R_\oplus$)\tablenotemark{d} &  $< 2.5 \pm 0.4$     \\
$M_p \,(M_J$)\tablenotemark{e} &  $< 1.4$ \\
$\theta_{*}$\tablenotemark{f} & $17^\circ \pm 1.5^\circ$\\
$\dot M_{\rm dust}$ (g s$^{-1}$)\tablenotemark{g}&   $\approx 2 \times 10^{11}$    \\
$\ell_2$ (leading tail only; units of $R_{\rm *}$)\tablenotemark{h} & $0.19-0.48$  \\
$a_{\rm grain}$ ($\mu$m)\tablenotemark{h} & $0.3 - 0.5$\\
Impact parameter, $b$\tablenotemark{h} &  $0.42-0.78$ \\
$\rho_1/\rho_2$ (two-tail model, $\ell_1 = \ell_2$)\tablenotemark{h} & $< 0.5$ 
\enddata
\tablenotetext{a}{Derived from the K2 and ground-based observations.}
\tablenotetext{b}{Based on the mass and radius of the host star given in Table \ref{tbl:stars} and Kepler's 3$^{\rm rd}$ law.}
\tablenotetext{c}{Derived from the K2 observations using a standard Mandel \& Agol (2002) fit to a hard-body transit.}
\tablenotetext{d}{Based on the shallowest K2 transits.}
\tablenotetext{e}{2-$\sigma$ limit based on the Keck RVs. }
\tablenotetext{f}{Estimate of the half angle subtended by the star at the position of the planet.} 
\tablenotetext{g}{Following the type of estimate made in Appendix D of Rappaport et al.~(2014)} 
\tablenotetext{h}{90\% confidence limits based on the models of section~\ref{sec:dusttail}. The leading and trailing tails are assumed to have exponential scale lengths and maximum optical thicknesses, $\ell_2$, $\ell_1$, $\rho_2$, and $\rho_1$, respectively. \vspace{0.1in}}

\vspace{0.1in}
\end{deluxetable}

\subsection{Spectral Studies}
\label{sec:spectraana}

\subsubsection{{\it NOT-FIES} spectrum}

An exploratory spectrum was obtained on 13 Feb 2015 with the FIbre-fed \'Echelle Spectrograph (Frandsen \& Lindberg~1999, Telting et al.~2014) mounted at the 2.56-m Nordic Optical Telescope of Roque de los Muchachos Observatory (La Palma, Spain). We used the 1.3$\arcsec$ \emph{Med-Res} fibre which provides a resolving power of $R=46000$ over the spectral range 3640--7360\,\AA. The FIES data revealed a single-lined spectrum. Although the low signal-to-noise ratio does not allow us to perform a reliable spectral analysis, a comparison with a grid of stellar templates from (Valdes et al.~2004) and (Bochanski et al.~2007) confirmed that the host is a cold dwarf star.

\subsubsection{{\it Keck-HIRES} spectra}

We also acquired five spectra of the host star with the Keck Telescope and HIRES spectrometer using the standard setup of the California Planet Search (CPS, Howard et al.~2010).  Over the course of four nights, (5-8 Feb 2015) we observed under clear skies and average seeing ($\sim$$1.2''$).  Exposure times of less than 5 minutes resulted in SNRs of 5-10 per pixel. Extensive scattered light and sky emission were unavoidable for exposures on 5 Feb 2015 due to the close proximity of the nearly full moon to the target star.All spectra were taken using the C2 decker, which is $0.87^{\prime \prime}$ wide  by $14^{\prime \prime}$ long, resulting in a spectral resolution of R=60,000. The slit was oriented to minimize the amount of contamination from the companion star.

We estimated the effective temperature \teff, surface gravity log\,g, iron abundance [Fe/H], and projected rotation velocity \vsini\ of the host star from the co-added HIRES spectrum, which has a S/N ratio of about 20 per pixel at 6000\,\AA. We used a modified version of the spectral analysis technique described in Gandolfi et al.~(2008), which is based on the use of stellar templates to simultaneously derive spectral type, luminosity class, and interstellar reddening from flux-calibrated, low-resolution spectra. We modified the code to fit the co-added HIRES spectrum to a grid of templates of M dwarfs recorded with the SOPHIE spectrograph (Bouchy et al.~2008). We retrieved from the SOPHIE archive\footnote{Available at \url{http://atlas.obs-hp.fr/sophie/}} the high-resolution ($R=75\,000$), high S/N ratio ($>$80) spectra of about 50 bright red dwarfs encompassing the spectral range K5--M3\,V. The stars were selected from the compilation of (Lepine et al.~2013). We downloaded spectra with no simultaneous thorium-argon observations, to avoid potential contamination from the calibration lamp.

The photospheric parameters of the template stars were homogeneously derived from the SOPHIE spectra using the procedure described in Maldonado et al.~(2015), which relies on the ratios of pseudo-equivalent widths of different spectral features. Unfortunately, the low S/N ratio prevented us from directly applying this technique to the co-added HIRES spectrum of the target.

Prior to the fitting procedure, the resolution of the template spectra was somewhat degraded to match that of the HIRES spectrograph ($R=60\,000$) -- by convolving the SOPHIE spectra with a Gaussian function mimicking the difference between the two instrument profiles. A corrective radial velocity shift was estimated by cross-correlating the observed and template spectra. We restricted the spectral range over which the fit is performed to 5500--6800\,\AA\ and masked out the regions containing telluric lines. We selected the 5 best fitting templates and adopted the weighted means of their spectroscopic parameters as the final estimates for the target star. We found that the target star has an effective temperature of \teff\,$=3780\pm90$\,K, surface gravity of log\,g\,$=4.65\pm0.12$ (log$_{10}$\,\cms2), and iron abundance of [Fe/H]\,$=0.05\pm0.08$~dex. We also set an upper limit of 1.5\,\kms\ on the projected rotation velocity \vsini\ by fitting the profile of several clean and unblended metal lines to the PHOENIX model spectrum (Husser et al.~2013) with the same parameters as the target star. Figure~\ref{HIRES_spectrum} shows the co-added HIRES spectrum in the spectal region around the H$\alpha$ line, along with the best fitting SOPHIE template.

\subsubsection{{\it IRTF-SpeX} spectrum}

A near-infrared spectrum of the star was also obtained using the updated SpeX (uSpeX) spectrograph (Rayner et al.~2003) on the NASA Infrared Telescope Facility (IRTF). SpeX observations were taken using the short cross-dispersed mode and the $0.3'' \times15''$ slit, which provides simultaneous coverage from 0.7 to 2.5\,$\mu$m at $R\simeq2000$.  The slit was aligned to capture both the target and companion spectrum. The pair was nodded between two positions along the slit to subsequently subtract the sky background. Ten spectra were taken following this pattern, which provided a final S/N of $\simeq100$ per resolving element in the $K$-band for the primary, and $\simeq$ 25 for the companion. The spectra were flat fielded, extracted, wavelength calibrated, and stacked using the \textit{SpeXTool} package (Cushing et al.~2004). An A0V-type star was observed immediately after the target, which was used to create a telluric correction using the \textit{xtellcor} package (Vacca et al.~2003). 

We analyzed the SpeX spectra obtained for both the bright and faint star, and both show strong atomic and weak CO absorption, as is expected for dwarf stars. Comparison with dwarf and giant NIR templates from the IRTF library (Rayner et al.~2009) rule out the possibility of either component being evolved. Metallicity was derived from the SpeX data using the procedures from Mann et al~(2013), who provide empirical relations between atomic features and M dwarf metallicity, calibrated using wide binaries. \teff\ was calculated using the empirical calibration from Mann et al.~(2013), which is based on stars with \teff\ determined from long-baseline optical interferometry (Boyajian et al.~2012). This analysis yielded a metallicity of 0.00$\pm$0.08 and a \teff\ of 3880$\pm$85~K for the primary; and a metallicity of 0.06$\pm$0.20 and a \teff\ of 3290$\pm$120~K for the companion. Both the metallicity and \teff\ determinations for the primary star are consistent with those derived from the analysis of the HIRES spectra.

\subsubsection{{\it UH88 SNIFS} spectra}

Finally, spectra of both K2-22 and its companion star were obtained with the SNIFS integral field spectrograph on the UH88 telescope on Mauna Kea during the night of UT 31 March 2015. The two stars were spatially resolved in the image cubes.  SNIFS spectra cover 3200-9700\,\AA\ with $R \approx 1000$, do not suffer from slit effects, and have been precisely calibrated by extensive observations of spectrophotometric standards (Lantz et al.~2004; Mann et al.~2011; Mann et al.~2013).  The wavelength coverage and resolution of SNIFS is more than adequate for measuring the strength of key molecular bands and atomic lines as indicators of $T_{\rm eff}$ and gravity for M dwarf stars. SNR $> 100$ was obtained for the primary.

The effective temperature of the target star was derived independently from the SNIFS spectrum by comparing it  to Dartmouth Stellar Evolution model predictions in a manner that has been calibrated to retrieve the bolometrically-determined temperatures of nearby stars with measured angular radii (Boyajian et al.~2012; Mann et al. 2013).  Radius and mass were then derived from empirical relations based on an expanded set of calibrator stars (Mann et al.~2015), with masses obtained from the Delfosse et al.~(2000) sample.  These yield $T_{\rm eff} = 3780 \pm 60$\,K, $R_* = 0.55 \pm 0.03 \, R_\odot$, $M_* = 0.57 \pm 0.06 \, M_\odot$, and $\log g = 4.72 \pm 0.05$, in quite good agreement with the other determinations of these parameters discussed above.

\begin{figure}
\begin{center}
\includegraphics[trim = {0cm 0cm 0cm 0cm}, clip, width=0.48 \textwidth]{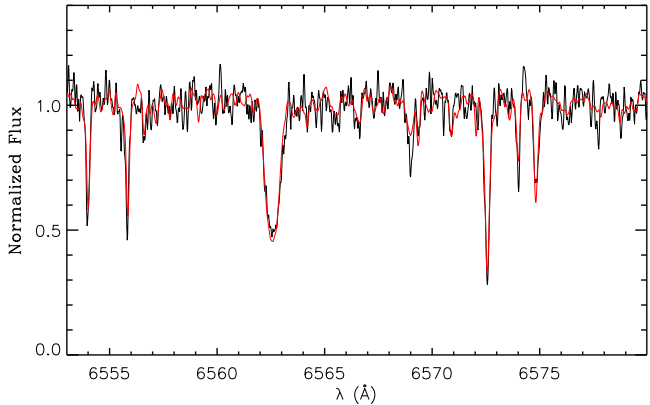}
\caption{HIRES co-added spectrum of K2-22 (black line) encompassing the H$\alpha$ line. The best fitting template spectrun is overplotted with a thick red line.}
\label{HIRES_spectrum}
\end{center}
\end{figure}

Adopting the mean [Fe/H] and \teff{} derived from SpeX and HIRES we derived $R_*$, $M_*$, and $L_*$ for the primary star, and we did the same with the SpeX parameters for the secondary star. To this end we utilized empirical \teff-[Fe/H]-$R_*$, \teff-[Fe/H]-$M_*$, and \teff-[Fe/H]-$L_*$ relations derived from the Mann et al.~(2015) sample. Accounting for errors in \teff, [Fe/H] and the Mann et al.~(2015) relations we computed physical parameters that are listed in Table~\ref{tbl:stars}. 

We also derived independent estimates of the distance to the stars, based on the colors and also on the spectroscopic parameters, reaching similar values with different methods and for both stars. We conclude that both stars are likely bound and at a distance of $225 \pm 50$ pc.

Constraints on the age of the system can be derived from the stellar rotation period using gyrochronology (see, e.g., Barnes 2007). It has been suggested that the classical relationships may not work for M dwarfs because of the large spread in rotation periods of stars in clusters (e.g., Reiners \& Mohanty~2012). However, the observed rotation period of the target star, 15.3 days (see Sect.~\ref{sec:aperture}), sits slightly above the main period distribution of M dwarfs in the Hyades and Praesepe clusters (Delorme et al.~2011), possibly indicating an older age. On the other hand, McQuillan et al.~(2013) finds a bimodal distribution of the rotation periods of field M dwarfs, likely associated with two populations with different median ages. K2-22 belongs to the shorter period (10-25 days), younger stellar population group. Both observations suggest that the age of the system could be between 1 and a few Gyrs.

The overall best-determined physical properties of the two stars, based on the spectroscopic observations and analyses, including $M_*$, $R_*$, $L_*$, $T_{\rm eff}$, $\log g$, metallicity, and distance, are given in Table \ref{tbl:stars}.

\section{Radial velocities}
\label{sec:RVs}

The five Keck HIRES spectra obtained can also be used to place constraints on the mass of the putative planet. The standard CPS pipeline is used to convert from raw spectra to two-dimensional spectra. Each of the three HIRES CCDs are independently reduced with flat fields, sky subtraction, and cosmic-ray removal. The pixel columns at each wavelength are then summed, resulting in flux as a function of wavelength for each pixel. Consistent wavelength solutions are insured by aligning a carefully chosen set of Thorium-Argon emission lines onto the same pixels at the beginning of each night's observations. 

The systemic radial velocity of each star is measured using the A-band and B-band telluric line features. Using the telluric lines as the wavelength fiducial, the relative placement of the stellar absorption lines is measured, and referenced to stars of known radial velocity (Chubak et al.~2012). These radial velocity measurements  are made relative the Earth's barycenter and are accurate to $\pm 0.3$ km/s. No RV variability in phase with the orbit of the target is detected, and the radial velocities have an rms of 0.3 km/s, compatible with the expected uncertainties.

We used these five Keck HIRES points to set a formal 1-$\sigma$ upper limit on the RV amplitude of the host star of 280 m/sec, yielding a $2-\sigma$ upper limit on the planet mass of 1.4 $M_J$.  This is not highly constraining in the context of a small rocky planet, but it does rule out non-planetary scenarios (assuming that the source of the photometric dips is the brighter star).

\section{GTC Multicolor observations}
\label{sec:GTC}

\subsection{Wavelength Dependent Transits}

Spectro-photometric observations for 3 complete transits were obtained with OSIRIS on the GTC (see also Sect.~\ref{sec:transits}).}  The GTC instrument OSIRIS consists of two CCD detectors with a field of view (FOV) of $7.8' \times 7.8'$ and a plate scale of $0.127''$ per pixel. For our observations, we used the 2 $\times$ 2 binning mode, a readout speed of 200 kHz with a gain of 0.95 e-/ADU and a readout noise of 4.5 e-. We used OSIRIS in its long-slit spectroscopic mode, selecting the grism R1000R which covers the spectral range of 520-1040 nm with a resolution of $R=1122$ at 751 nm. The observations and results we present here were taken using a custom built slit of $12''$ in width, with the target and a comparison star both located in the slit. The use of a wider slit has the advantage of reducing the possible systematic effects that can be introduced by light losses due to changes in seeing and/or imperfect telescope tracking  (Murgas et al. 2014).

During the first transit observed with the GTC (29 Jan) we took as a reference a very close comparison star to the east of the target. However, for the two last transits (13 and 14 Feb) we took a reference star with a similar brightness to K2-22 and located at a distance of 2.7$'$ from the target. The position angle of the reference star with respect to the target was  $-79.2^\circ$. The two stars were positioned equidistantly from the optical axis, close to the center of CCD\#1, while CCD\#2 was turned off to avoid crosstalk.

The basic data reduction of the GTC transits was performed using standard procedures. The bias and flat field images were produced using the Image Reduction and Analysis Facility (IRAF\footnote{IRAF is distributed by the National Optical Astronomy Observatory, which is operated by the Association of Universities for Research in Astronomy (AURA) under a cooperative agreement with the National Science Foundation.}) and were used to correct the images before the extraction of the spectra. The extraction and wavelength calibration was made using a PyRAF\footnote{Python environment for IRAF} script written for GTC@OSIRIS long-slit data. This script automated some of the steps to produce the spectra such as: extraction of each spectrum, extraction of the corresponding calibration arc, and wavelength calibration (using the HgAr, Xe, Ne lamps provided for the observations).  All spectra were aligned to the first spectrum of the series to correct for possible shifts in the pixel/wavelength solution during the observations caused by flexures of the instrument. Several apertures were tested during the reduction process, and the one that delivered the best results in terms of low scatter (measured in rms) in the points outside of transit for the white light curve was selected. The results presented here were obtained using apertures of 28, 40, and 44 pixels in width for the three transit observations, respectively. Final spectra were not corrected for instrumental response nor were they flux calibrated. Fig.~\ref{fig:gtc1} shows the extracted spectrum of K2-22.

\begin{figure}[h]
\begin{center}
\includegraphics[width=0.99 \columnwidth]{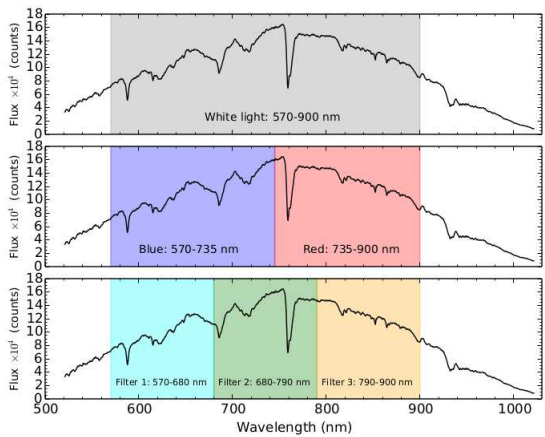} 
\caption{Visual explanation of our different choices for splitting the GTC observations into several different broad color bands. These are centered at  652.5 nm and 820 nm for the `red' and `blue' colors, and 625 nm, 735 nm, and 850 nm respectively, for the 3-color bands.}
\label{fig:gtc1}
\end{center}
\end{figure}

The universal time of data acquisition was obtained using the recorded headers of the spectra indicating the opening and closing time of the shutter in order to compute the time of mid exposure. We then used the code written by Eastman et al.~(2010)\footnote{http://astroutils.astronomy.ohio-state.edu/time/} to compute the BJD time using the mid-exposure time for each of the spectra 
to produce the light curves analyzed here. 

We constructed light curves over several spectral ranges to search for color dependencies of the transit shape and depth. In the case of the white light curve, the flux was integrated over almost the entire wavelength range of the observed spectra, between 570 and 900 nm, but avoiding the blue and red ends of the spectra where the SNR is lower. This step is particularly important, because observations at redder wavelengths suffer from fringing\footnote{http://www.gtc.iac.es/instruments/osiris/osiris.php\#Fringing} whereas using the data at bluer wavelengths not only did not improve the quality of the light curves, but increased the occurrence of outliers. Three narrower color light curves were created from the raw spectra for each star, by integrating all counts over the spectral ranges 570-680 nm, 680-790 nm, and 790-900 nm (see Fig.~\ref{fig:gtc1}). The final light curves used in the analyses were created by dividing the K2-22 light curves by the comparison star light curves. The scatter (std) for the relative white light curves can be found in Table~\ref{tbl:log}.

The white light  transit profiles of the three separate GTC transits were shown earlier in Fig.~\ref{fig:groundtrans}. It is obvious that the depth of the transits change from one night to the next, confirming the rapidly varying nature of this object. Note that between transits 2 and 3, only 27.44 hours had passed and the transit depth nearly doubled. 

\begin{figure*}[ht]
\begin{center}
\includegraphics[width=0.96 \textwidth]{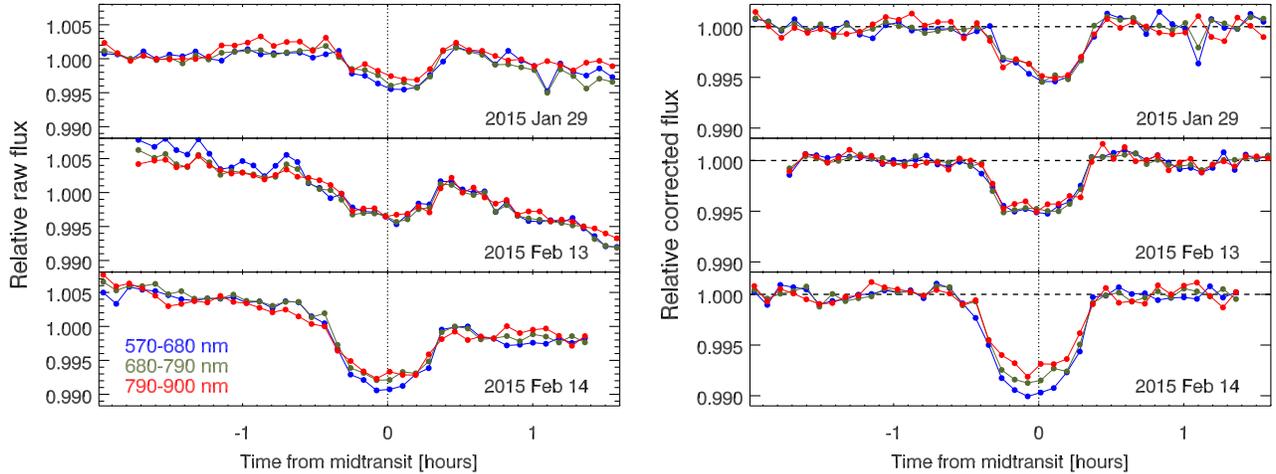}
\caption{GTC observations split among three color bands. On the left panels we show the raw light curves, which have been median-normalized. The observations shown on the right hand side have been corrected for time, airmass and seeing effects. The first two transits were shallower, and depth variations are hard to notice, but during the third night, when the white light transit is deepest, color variations are clearly observed.}
\label{fig:gtc3}
\end{center}
\end{figure*}

To further investigate the nature of these changes in the
GTC transit curves, in Fig.~\ref{fig:gtc3} we plot the three color light
curves for each transit separately. Transit 3 (bottom panel) is
the deepest and shows a clearly increasing transit depth towards
the blue. The SNR is lower in the other two sets of
color light curves and the transits are shallower, and thus there
is no clear trend in transit depth with wavelength. In order to
interpret quantitatively the color dependence of the depths in
the third GTC transit,  we fit our standard three-segment transit
profile model to the data from each of the nights, with three different transit depths to 
evaluate how the transit depth changes with color. We also fit each transit with five additional parameters, to construct a polynomial that reaches second order in time from mid-transit and it linearly depends on the mean-subtracted airmass and full-width half maximum of the images of the host star. The total number of parameters is 21 for each observation, with a total 120, 132 and 114 data points in each of the three nights. After finding the best-fit model, we set the error of the flux measurements of each light curve to provide a best-fit standard $\chi^2$ value equal to the number of degrees of freedom. We then took correlated noise into account computing the $\beta$ factors for each light curve and each night (see section~\ref{sec:transits}), and multiplying the errors by their corresponding $\beta$ factor. We use an MCMC routine to obtain posterior distributions for the transit depths, which are marginalized respect to all the other 18 model parameters (see Table~\ref{tbl:depths}).

\begin{deluxetable}{lcccc}

\tablecaption{Multicolor observations summary\label{tbl:depths}}
\tablewidth{0pt}

\tablehead{
\colhead{Date} &
\colhead{$\delta_{\rm 570-680 nm}$}  & 
\colhead{$\delta_{\rm 680-790 nm}$ }  & 
\colhead{$\delta_{\rm 790-900 nm}$}  & 
\colhead{$\alpha$}  
}

\startdata
Jan 29  &     $0.50 \pm 0.06$\% & $0.47 \pm 0.05$\%  & $0.46 \pm 0.05$\% & $0.13 \pm 0.55$ \\
Feb 14  &     $0.48 \pm 0.04$\% & $0.49 \pm 0.03$\%  & $0.44 \pm 0.03$\% & $0.11\pm 0.37$ \\
Feb 15  &     $0.95 \pm 0.04$\% & $0.83 \pm 0.03$\%  & $0.72 \pm 0.04$\% & $0.83 \pm 0.23$ 
\enddata

\tablenotetext{}{These transit depths are measured from the data alone, and do not take into account contamination from the faint companion or airmass corrections.   The calculated Angstr\"om exponent, $\alpha$, is, however, corrected for the small dilution factor of the faint companion star.}
\end{deluxetable}

From the marginalized posterior distributions we compute a quantity called the ``Angstr\"om exponent'', $\alpha$, which is defined as $-d\ln \sigma /d\ln \lambda$ , where $\sigma$  is the effective extinction cross section. Here we take the transit depth (on each night) at wavelength, $\lambda$, to be proportional to the effective cross section at $\lambda$ (under the assumption that the dust tail is optically thin). In order to use the transit depths, we first multiply each individual transit depth by $(1+D_i)$ to apply a dilution correction, where each $D_i$ is the flux ratio between the faint companion and the host star evaluated at the flux weighted center of each selected band.  These $D_i$ values are  obtained from our SED models (see Sect.~\ref{sec:image}), and  have values of 4.4\%, 6.6\% and 8.9\% with increasing wavelength, with uncertainties of 0.2\%. We then find $\alpha = 0.83 \pm 0.23$, for the transit observed on February 15. For the other two transits, the values of $\alpha$ are $0.11 \pm 0.37$ and $0.13 \pm 0.55$, both slightly positive, but not 
statistically significantly different from zero.  We made no corrections
for the wavelength-dependent limb darkening properties, but utilized
the quadratic limb darkening coefficients of Claret \& Bloemen (2011)
to estimate that the values of $\alpha$ would be {\em lowered} by between
0.15 and 0.02, for an equatorial transit (which is unlikely) and an impact 
parameter of $b = 0.7$, respectively. For higher impact parameters,
up to 0.85 (see Table \ref{tbl:planet}), the value of $\alpha$ could 
actually be {\em raised} by up to 0.15, and become even more 
significantly different from zero.

\subsection{Interpretation in Terms of Dust}

We interpret the value of the Angstr\"{o}m exponents in terms of Mie scattering (for spherical dielectric dust particles) with a variety of different compositions.  We computed the Angstr\"{o}m exponent, $\alpha$, over the range 630-840 nm for a set of power-law distributions for the dust particle sizes, with $dN/da \propto a^{-\Gamma}$.  In order to guarantee that the total cross section converges, we also need to specify a maximum grain size, $a_{\rm max}$.  For the sake of specificity we adopted an illustrative dust composition of corundum, but the conclusions we draw are the same for a number of other common refractory materials, and in fact, for any material with real and imaginary indices of refraction of $n \simeq 1.6$ and $0.001 \lesssim k \lesssim 0.03$.  We plot the computed Angstr\"{o}m exponent in Fig.~\ref{fig:angexp} as a function of $a_{\rm max}$ for five different power-law exponents, $\Gamma$.  As can be seen from the figure, values of $\alpha$ in the range of $\sim$$0-1$, as indicated in Table \ref{tbl:depths}, correspond to non-steep power-law indices of $\Gamma \simeq 1-3$ and maximum particle sizes of $\sim$$0.4-0.7\,\mu$m.  In turn, these correspond to ``effective particle sizes'' of 0.2 $\mu$m to 0.4 $\mu$m, where the effective radius is the average grain radius weighted by both the size distribution and the cross section, i.e., 

\begin{equation}
a_{\rm eff} =\int_0^{\rm a_{max}} a^{1-\Gamma} \, \sigma(a,\lambda) \,\,da \,\,/ \int_0^{\rm a_{max}} a^{-\Gamma} \,
\sigma(a,\lambda) \,\,da
\end{equation}

where $\sigma(a,\lambda)$ is the wavelength dependent Mie extinction cross section for a particle of radius $a$.

\begin{figure}[h]
\begin{center}
\includegraphics[width=0.48 \textwidth]{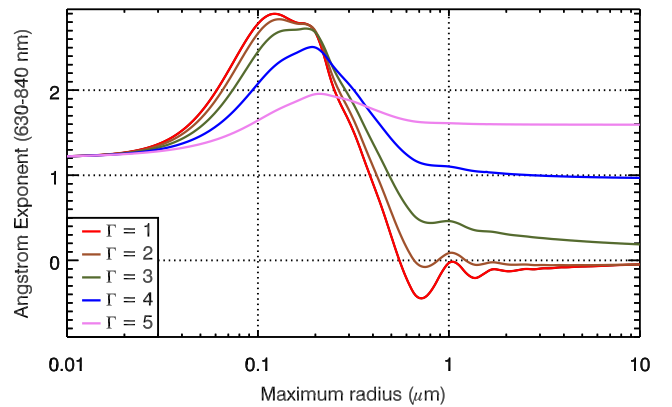}
\caption{The Angstr\"{o}m exponent computed from Mie scattering over the wavelength range $0.63-0.84\,\mu$m as a function of the maximum particle size, for five different indices, $\Gamma$, of the power-law grain size distribution, i.e., $dN/da \propto a^{-\Gamma}$. We adopted an illustrative particle material of corundum.}
\label{fig:angexp}
\end{center}
\end{figure}

{\section{Why The Brighter Star is the Source of the Transits}
\label{sec:brightstar}

Here we summarize why we are confident that it is the brighter star that is the source of the observed transits.  First, the fainter star is redder than the bright star (see Sect.~\ref{sec:image}); in particular the contribution from the faint star to the total flux increases by a factor of 2 from r to i band and a factor of 4 from r to z band. If we assume that the {\em faint} star is the host, and assume an achromatic mean transit depth, the changes in flux ratio should translate into correspondingly larger transit depths toward the red due to the dilution effect. In section~\ref{sec:GTC} we showed simultaneous GTC transit observations in bands similar to r and a combination of the i and z bands. The transit depth, at least on one occasion, is actually considerably {\em shallower} in the latter redder band, strongly suggesting that the bright star is the one being transited.  

From another line of argument, we note that the transit depths are sometimes as large as 1.3\%, and if the fainter star (5\% of the flux in g-band) were the source of the transits, it would have to be attenuated by 24\% of its flux.  Since the transits are highly variable in depth, and the transit profile is not that due to a conventional hard-body transit, it is almost certainly due to a `soft' attenuator such as a dust tail.  But, the fainter star has a radius of 0.3 $R_\odot$, and therefore an occulting dust tail would have to either cover the entire star with an optical depth, $\tau$, of $\sim$0.25, or cover $\sim$25\% of the star (i.e., a tail thickness of $\sim$0.08 $R_\odot$) with $\tau \simeq 1$.  Simulations of the dust tail in KIC 1255b (Rappaport et al.~2012) and this object (see Sect.~\ref{sec:headtail}) indicate that the vertical thickness of the dust tail is only $\sim$0.03 $R_\odot$. This would make the scenario of a dust tail covering over 1/4 of the area of the fainter star highly unlikely.

\section{The disintegrating planet hypothesis}
\label{sec:dusttail}

\subsection{Evidence for a Disintegrating Planet}
\label{sec:disinteg}

When taken together, all these observational results point clearly toward a planet in a 9-hour orbit that is disintegrating via the emission of dusty effluents.  The lines of evidence pointing in this direction include: (1) erratically and highly variable transit depths (see Sect.~\ref{sec:variability}); (2) transit profile shapes from ground-based
observations that are likely variable (though with lesser confidence that the depth changes; see Sects.~\ref{sec:transits} and \ref{sec:GTC}); and (3) an average transit profile from the K2 data that exhibits clear evidence for a post-transit ``bump'' and also weaker evidence for a pre-transit ``bump'' (Sect.~\ref{sec:diverge}).  The first of these is highly reminiscent of the disintegrating planet KIC 1255b (Rappaport et al.~2012), while the variable transit shapes are also detected in KIC 1255b from ground-based studies (R.~Alonso et al.~private communication; Bochinski et al.~2015).  The transit profile of K2-22b, with a post-transit ``bump'', is different from the transit profiles of KIC 1255b and KOI 2700b which show a post-transit depression as opposed to a post-transit bump.  The first two of the above listed features point to obscuration by dusty effluents coming from a planet, while the third property needs to be explained in this same context.  In the following sections we explore the significance and the interpretation of the transit profile.

\subsection{Quantitative Model for the Leading Dust Tail}
\label{sec:headtail}

A dust tail emanating from a planet, as inferred in the cases of KIC 1255b and KOI 2700b, trails the planet as is illustrated  in Fig.~6 of Rappaport et al.~(2012; see also the middle panel of our Fig.~\ref{fig:dust_tails}). Such dust tails are the way they would be seen in the reference frame of the planet, and note that the motion of the planet is implicitly in the opposite direction from the tail. In this case we would say that the tail ``trails the planet''.  The reason for this is that the radiation pressure acting on the dust forces it into an eccentric orbit with its periastron located at the point where the particle was released.  This orbit has a larger semimajor axis than that of the planet (see Appendix B of Rappaport et al.~2014).  In turn, the larger orbit has a lower orbital frequency, and the particles appear to trail behind the planet in a comet-like tail.  

An observer viewing this system is effectively moving counterclockwise.  Thus, the ingress to the transit is sharp as the ``comet head'' moves onto the stellar disk (see Fig.~6 in Rappaport et al.~2012).  The trailing tail would lead to a depression upon egress as the tail slowly moves off of the stellar disk.  In that case, the pre-transit ``bump'' is caused by forward scattering by the densest regions of the dust which have not yet reached the stellar disk. This is the case we believe we see in KIC 1255b (Rappaport et al.~2012; Brogi et al.~2012; Budaj 2013; van Werkhoven et al.~2014).  If the tail of the planet is sufficiently short (i.e., compared to the radius of the host star) then there would be both a pre-transit and a post-transit ``bump'', the latter of which would dominate over the relatively shallow post-transit depression (see also Budaj 2013).  In this case, the pre-transit bump would be somewhat larger due to the asymmetry in the direction of the tail.  

A logical first guess as to how to produce a post-transit ``bump'' on the transit curve would be to reverse the direction of the comet-like tail.  However, as we have seen, substantial radiation pressure forces inevitably lead to a trailing tail.  Then, the question becomes how to produce a ``leading dust tail'' to the planet.  One way to have dusty material lead the planet, i.e., moving faster, would be to have it overflow its Roche lobe (or, Hill sphere of influence) and fall in toward the host star.  At first consideration this doesn't seem to work since a rocky planet in a 9-hour orbit will not be close to filling its Roche lobe.  Rappaport et al.~(2013) showed that the critical density for Roche-lobe overflow is largely a function of its orbital period, and is nearly independent of the properties of the host star.  In particular, Eqn.~(5) in Rappaport et al.~(2013) suggests that critical density for a planet to be filling its Roche lobe is
\begin{equation}
\rho_{\rm crit} \simeq \left(\frac{11.3 \, {\rm hr}}{P_{\rm orb}}\right)^2 ~{\rm g/cm}^3 \simeq 1.5 ~{\rm g/cm}^3 
\end{equation}
where the right-hand value is for a 9-hour planet.  It seems very unlikely that a planet with this low a mean density would be disintegrating via dusty effluents.  Turning the problem around, we can ask what fraction of the Roche-lobe radius would be occupied by a planet with a mean density in the range of 5-8 g/cc, which might be more appropriate for a dust emitter (see, e.g., Rappaport et al.~2013).  It is then evident that for densities which are $\sim$$3-5$ times higher than their critical densities, their radii are not even factors of $\sim$2 times smaller than their Roche lobes.  

Therefore, we come to the conclusion that planets with rocky compositions in a 9-hour orbit are underfilling their Roche lobes by only a factor of $\sim$2.  This is more than sufficient to prevent the planet from directly overflowing its Roche lobe. However, the potential difference between the planet's surface and the Roche lobe is about half the potential difference to infinity.  Thus, if the mechanism that drives off the dust or the heavy metal vapors that condense into dust, e.g., via a Parker-type wind (Rappaport et al.~2012; Perez-Becker \& Chiang 2013) imparts the full escape velocity of the planet or more, then the material can certainly reach the surface of its Roche lobe.

\begin{figure}
\begin{center}
\includegraphics[width=0.99 \columnwidth]{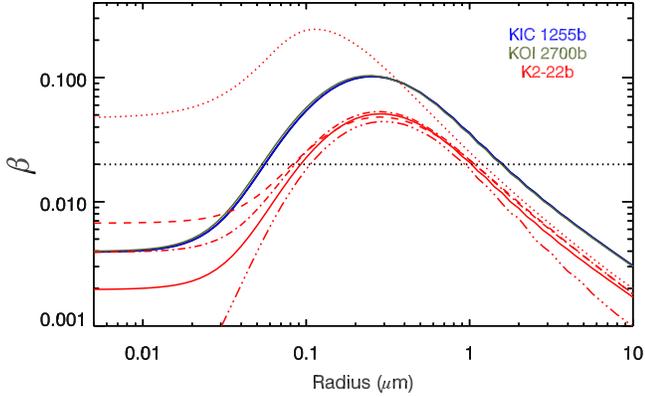} 
 \caption{Calculated values of $\beta$, the ratio of radiation pressure to gravitational forces, as a function of particle size. The blue and green curves (nearly superposed) are for KIC 1255b and KOI 2700b, respectively.  The red curves are values of $\beta$ for EPIC 201673175, for different materials.  For KIC 1255b and KOI 2700b we adopted $n = 1.65$ and $k = 0.01$ for the real and imaginary parts of the index of refraction.  The solid red curve is for the same indices.  The  dot-dashed curve is for the same real index but with $k = 0.02$, while the dotted, dashed, and dot-dot-dashed curves are specifically for iron, corundum and forsterite, respectively.  The horizontal black dotted line indicates the value of $\beta = 0.02$ below which most dust particles would go into a leading tail.}
 \label{fig:betas}
 \end{center}
 \end{figure}

If the material reaching the Roche surface feels a substantial radiation pressure, it will still be blown back into a comet-like tail.  By ``substantial'' we mean that $\beta$, the ratio of radiation pressure forces to gravitational forces, exceeds a certain value such as $\beta \gtrsim 0.05$.  For sufficiently small values of $\beta$, however, particles initially directed toward the host star (which seems reasonable since the gas or dust emission should commence on the heated hemisphere) will largely fall toward the host star until the grains are pushed into orbit by coriolis forces.  These particles will form a ``leading tail'' in the sense that the motion will be in front of the planet.

\begin{figure}
\begin{center}
\includegraphics[width=0.99 \columnwidth]{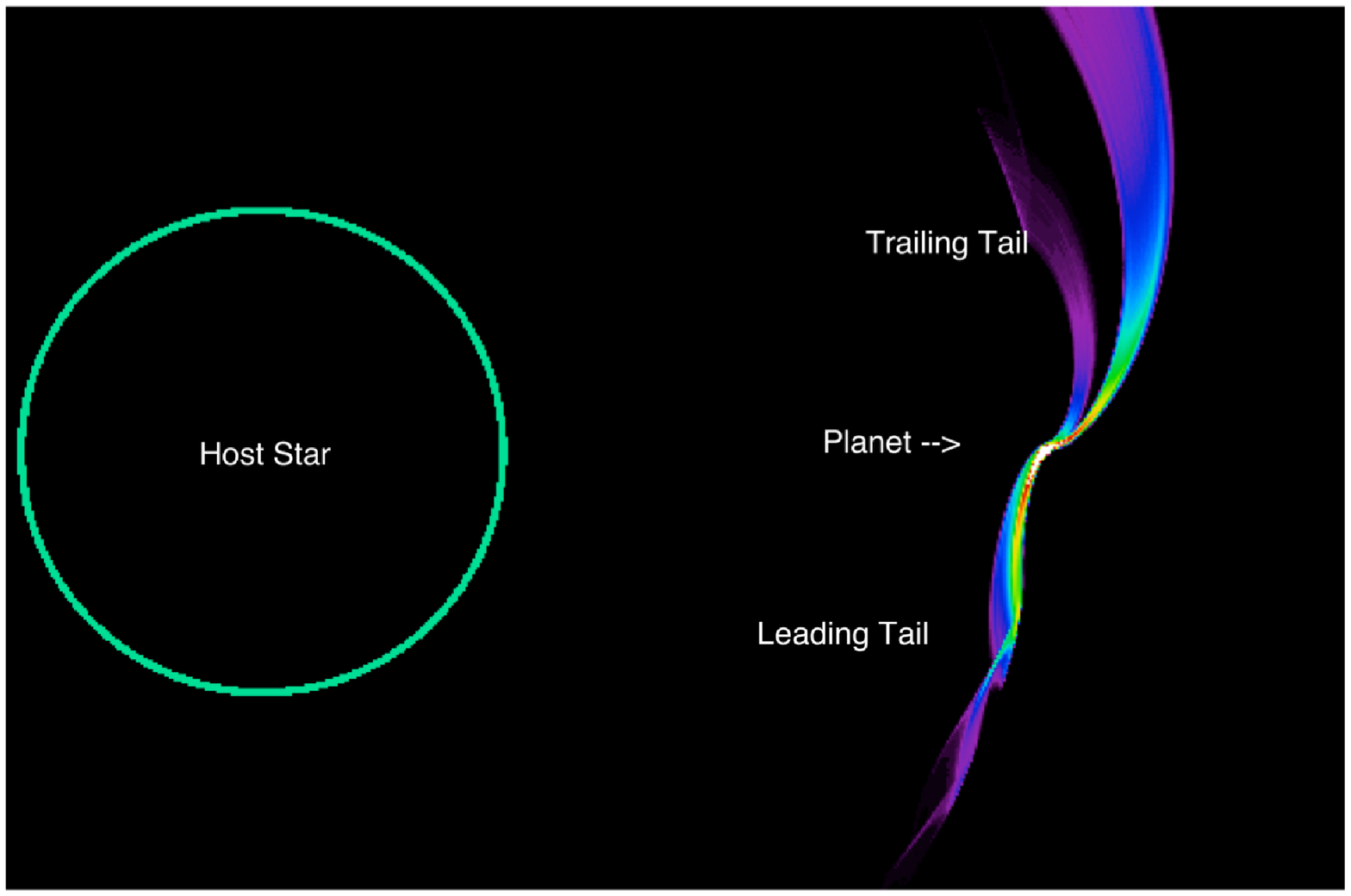} \vglue0.05cm
\includegraphics[width=0.99 \columnwidth]{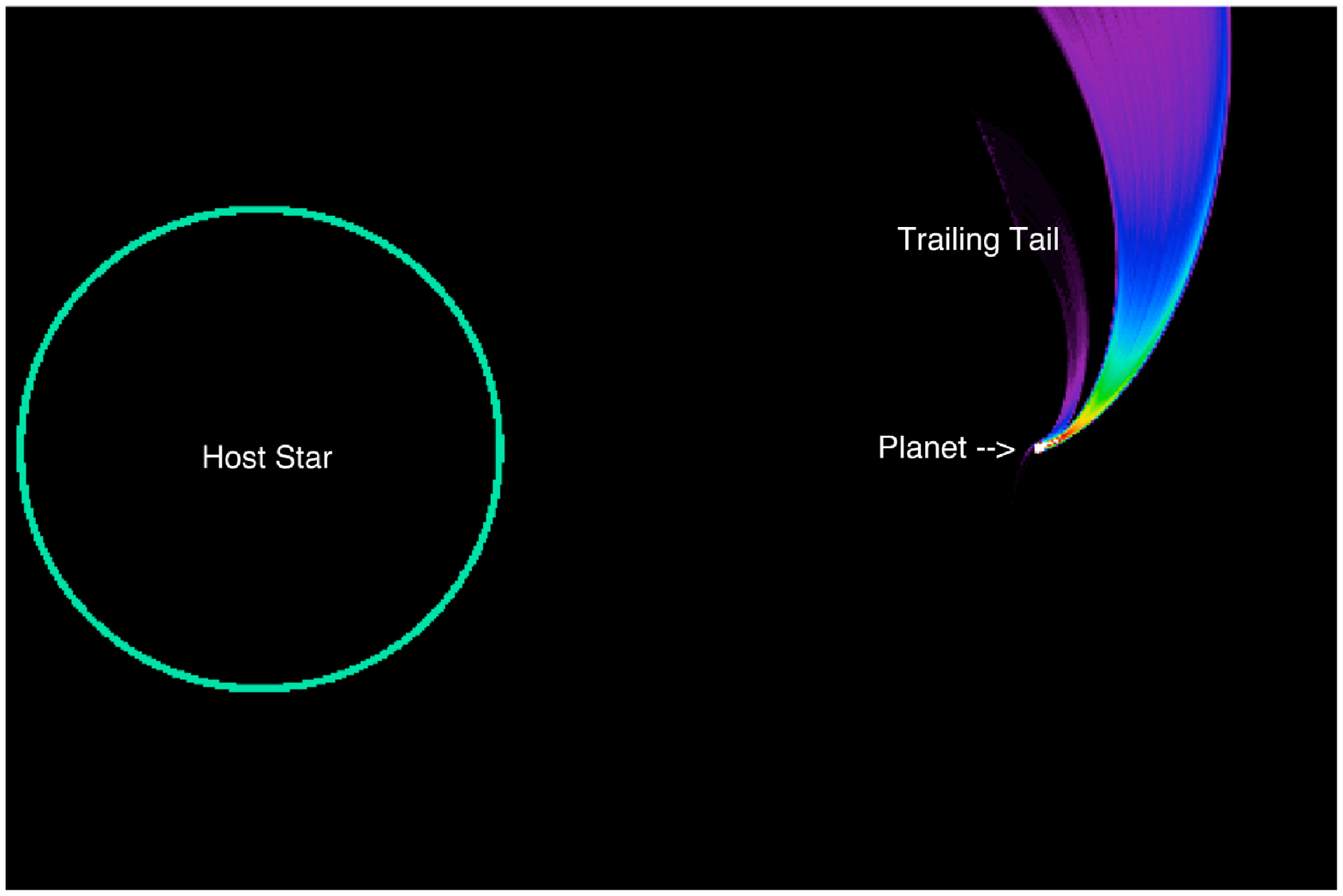} 
\includegraphics[width=0.99 \columnwidth]{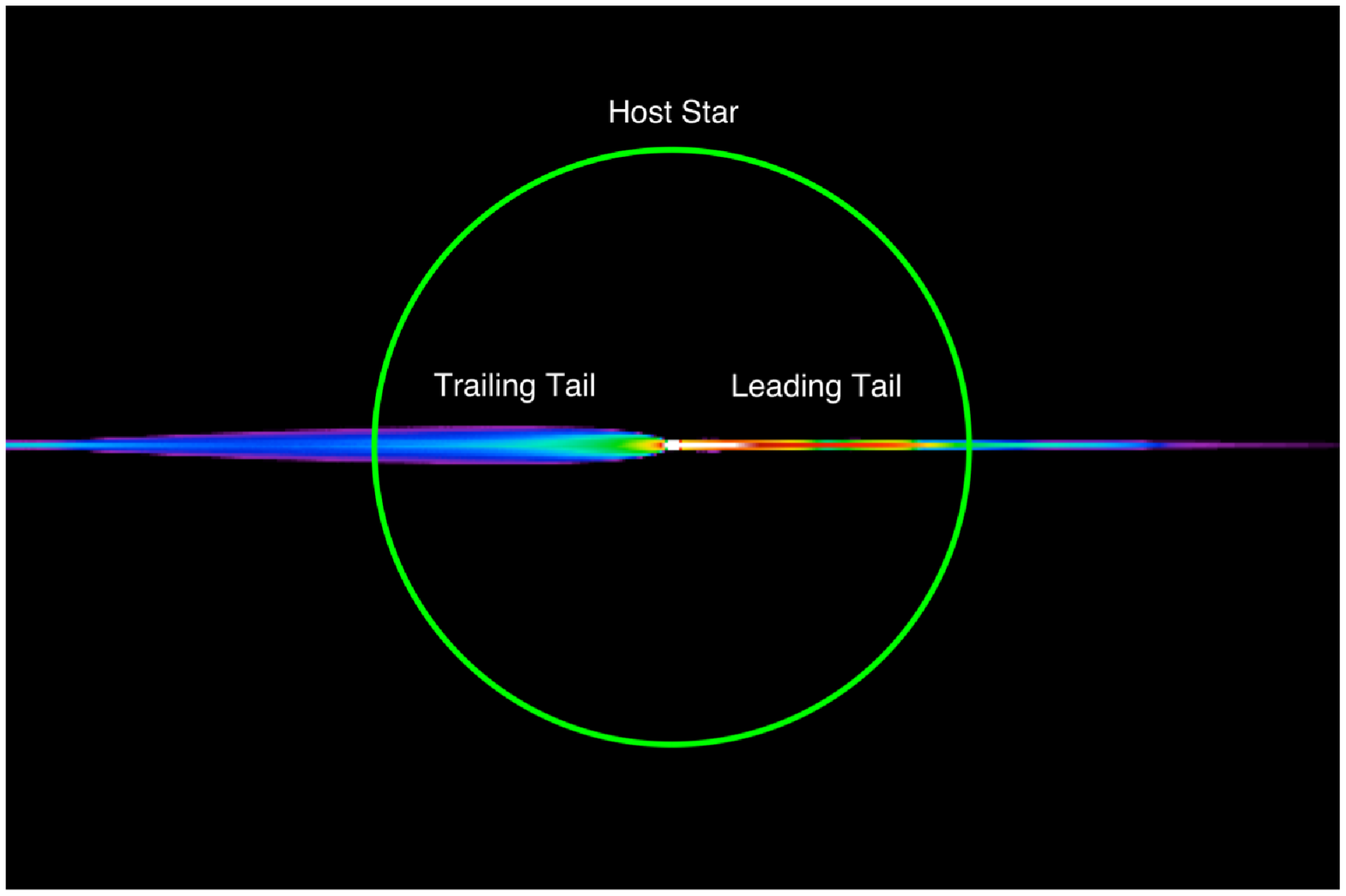} 
 \caption{Simulated dust tails.  The dust particles are launched with somewhat more than the escape speed from the planet within a 30$^\circ$ cone centered on the direction of the host star.  The particle sizes are chosen via Monte Carlo means from a power-law distribution, and the value of $\beta$ for each particle is calculated from the system parameters (see text). There are 50,000 particles in each simulation, and each one is assumed to have an exponentially decaying cross section (see App.~C of Rappaport et al.~2014) with a time constant of $10^4$ s.  Only radiation and gravitational forces are included.  The color coding is proportional to the logarithm of the dust particle density with white-red the largest to blue-purple the lowest.  The dust tails are shown in the rest frame of the orbiting planet (implicitly moving downward in the frame).  The top panel is a view of the orbit and dust tail from the orbital pole, and clearly shows both a leading and a trailing tail. The bottom panel is a view from the orbital plane as the dust emitting planet crosses the disk of the host star.  The middle panel results from arbitrarily multiplying each calculated value of $\beta$ by a factor of 4, the net effect of which is to eliminate the leading tail.}
 \label{fig:dust_tails}
 \end{center}
 \end{figure}

\begin{figure}
\begin{center}
\includegraphics[width=0.95 \columnwidth]{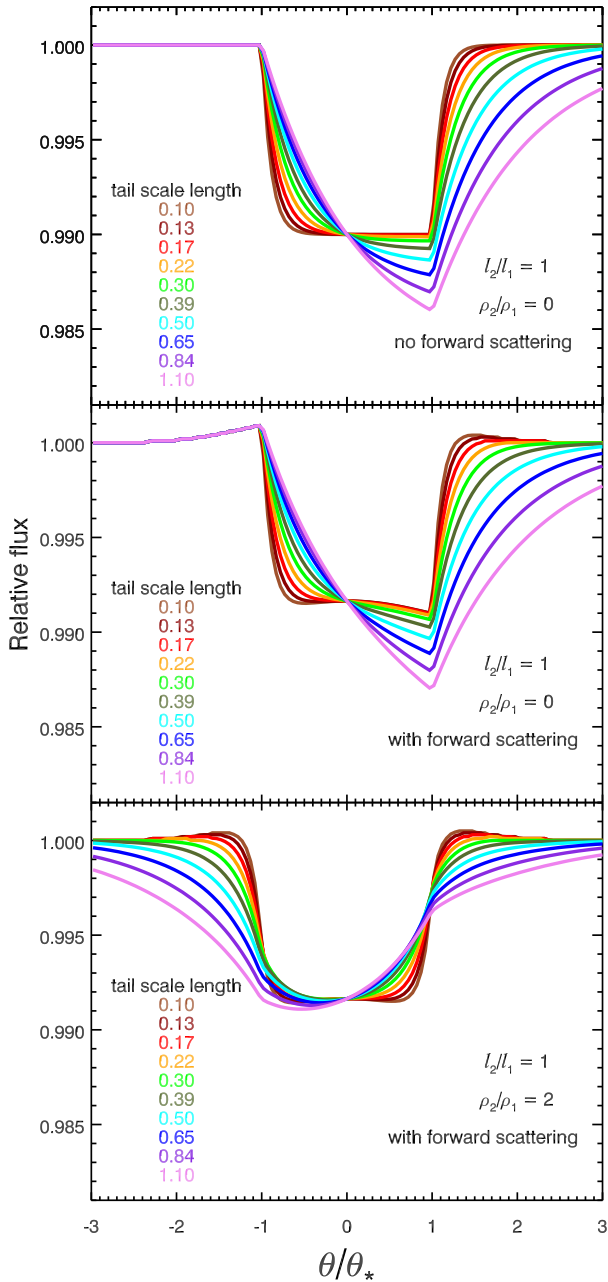} 
 \caption{ Schematic transit profiles for a range of dust tail parameters. $\theta$ is the orbital phase of the planet, and $\theta_{*}$ is the half angle subtended by the host star as seen from the planet.  {\it Top panel}: Trailing dust tail only.  The exponential tail length, $\ell$, is color coded and is expressed in units of the radius of the host star.  No forward scattering is included.  Note how the shorter the tail length, the smaller is the post-transit depression. {\it Middle panel}:  Same as for the top panel, but in this case a contribution from forward scattering has been included.  There is now a noticeable pre-transit bump due to the forward scattering, but only for the case of short dust tails is there a barely perceptible post-transit bump; it is typically suppressed by the post-transit depression.  {\it Bottom panel}: Here, there is both a trailing and a leading dust tail.  The exponential scale lengths, $\ell$, of both dust tails are assumed to be the same, but the leading tail is taken to have twice the dust density, $\rho$, at every corresponding angular distance, as the trailing tail.  For long tail lengths there are no pre- or post-transit bumps due to forward scattering, as they are pulled below the discernible level by the corresponding depressions caused by direct attenuation from the tails.  By contrast, for the case of short tail lengths, i.e., $\lesssim 0.2 \, R_{\rm star}$, there is both a pre- and a post-transit bump, with the latter being slightly larger.}
 \label{fig:profiles}
 \end{center}
 \end{figure}

The values of $\beta$ for the particles are computed according to 
\begin{equation}
\label{eqn:beta1}
\beta \simeq \frac{L_{\rm *} \sigma}{4 \pi c G M_{\rm *}\mu} \equiv \frac{3 \,L_{\rm *}  \, \langle \tilde{\sigma} (a,\lambda) \rangle}{16 \pi c G M_{\rm *} \rho a}
\end{equation}
where $L_{\rm *}$ and $M_{\rm *}$ are the luminosity and mass of the host star, $a$ is the particle radius, $\sigma$ is the particle cross section at wavelength $\lambda$, $\langle \tilde{\sigma} \rangle$ is the dimensionless cross section in units of $\pi a^2$ averaged over the stellar spectrum (see, e.g., Kimura et al.~2002), and $\mu$ and $\rho$ are the mass and material density of the dust grains.  Since, for low-mass main-sequence stars the luminosity scales roughly as $M_{\rm *}^4$, we find that for a fixed material in the grains, $\beta$ scales as 
\begin{equation}
\label{eqn:beta2}
\beta~ \propto~ \frac{M_{\rm *}^3 \,\langle \tilde{\sigma}(a,\lambda) \rangle }{a}
\end{equation}
We evaluate the dimensionless spectrum-averaged cross section with a Mie scattering code (Bohren \& Huffman 1983) for different assumed material indices of refraction (see Croll et al.~2014).   Plots of $\beta(a)$ for several different assumed compositions are shown in Fig.~\ref{fig:betas}.  They are also compared against representative $\beta(a)$ curves for KIC 1255b and KOI 2700b.  The latter two systems have $\beta(a)$ curves that are essentially a factor of 2 higher than for K2-22b, largely due to the lower luminosity per unit mass of the latter host star.  

Expression (\ref{eqn:beta2}) raises the interesting prospect that for low mass stars, the value of $\beta$ will be sufficiently small that outflowing particles passing through their Roche lobes will be immune to radiation forces and act much in the same way as Roche-lobe overflowing material.

We next carried out a large number of simulations of dust particles ejected from a planet and initially moving in various directions with different speeds.  For most of our simulations, we took the direction of the particles to be uniformly distributed within a cone of 30$^\circ$ radius and centered in the direction of the host star.  The velocities were somewhat arbitrarily taken to be one half the escape speed from the surface of the planet's Roche lobe to infinity in the absence of the host star, i.e., $\sqrt{GM_{\rm p}/2R_L}$ where $M_{\rm p}$ is the planet's mass and $R_L$ is its Roche-lobe radius.  This is the minimum that can be contemplated for a Parker-wind type outflow.  For each dust grain, a radius, $a$, is chosen from an assumed particle size distribution via Monte Carlo methods.  We considered a power-law differential size distribution for the dust grains with a slope of $-2$, as illustrative, with the maximum and minimum sizes in this distribution taken to be 1 $\mu$m and 1/20 $\mu$m, respectively.   The luminosity of the host star is taken to be that of a main-sequence M-K star of 0.6 $M_\odot$ (see Table \ref{tbl:stars}).  For purposes of the numerical calculations of the particle dynamics only, we used a simple analytic fit to the plots shown in Fig.~\ref{fig:betas} of the form:  
\begin{equation}
\label{eqn:beta3}
\beta ~\propto ~\frac{\langle \tilde{\sigma}(a,\lambda)\rangle}{a} ~\propto ~ \frac{c_1+c_2 a^3}{1+c_3 a^4}
\end{equation}
where the $c$'s are constants to be fit, and which depend on $L_*$, $M_*$, and the dust composition.

The results of our particle dynamics simulations are shown in Fig.~\ref{fig:dust_tails}.  The top panel represents the trajectories of all particles regardless of their size or corresponding value of $\beta$. The dust density is assumed to decay exponentially in time, with a time constant of $10^4$ sec.  Note that there is a forward moving ``leading tail'' in addition to a bifurcated trailing tail.  The inner of the two trailing tails arises from particles with small values of $\beta$ that are launched with a substantial velocity component in the forward direction.  In the bottom panel of Fig.~\ref{fig:dust_tails} we see the two tails from the perspective of an observer viewing an equatorial transit; there are far more particles in the leading tail.  For comparison, we show in the middle panel a case where the computed value of $\beta$ was arbitrarily multiplied by a factor of 4; all other parameters remained the same.  Note that the result is a purely trailing dust tail as one would have in a solar system comet.  This latter exercise ensures that all the particles have a $\beta$ that exceeds a critical value of $\sim$0.02, guaranteeing that the particles will go into a trailing tail.  The bottom line is that dust-emitting planets around luminous stars should have predominantly trailing tails while the reverse is true for low-luminosity stars.

In all of these calculations, we have ignored possible ram pressure forces on the dust grains due to a stellar wind from the host star.  For a justification of why this may be a good approximation, see Appendix A of Rappaport et al.~(2014).

\vspace{0.1cm}

\subsection{Model Transits}

\subsubsection{Idealized Transit Models}

The transit profiles expected for a planet with a trailing comet-like tail are already sufficiently complicated compared to a conventional hard-body planet transit.  They include such issues as the attenuation caused by the (unknown) absorption profile, typically characterized by an exponential scale length\footnote{For a discussion of why the dust tail may fall off approximately exponentially, including the effects of dust sublimation on a timescale of hours, see Appendix C of Rappaport et al.~(2014).}; possible forward scattering which is influenced by the grain size and composition (which may be changing along the tail as different materials sublimate at different rates); convolution of the scattering phase function with the finite angular profile of the host star; and another convolution with the density profile of the tail.  When adding the possibility of both a trailing and leading tail, as we have postulated in this work, the situation quickly becomes even more complicated. In an attempt to keep the numbers of free parameters to a minimum, we have adopted the following seven-parameter model: an exponential attenuation profile of the tail in both directions, each with its own scale length, $\ell$; a relative density between the two tails, $\rho_1/\rho_2$; the size of the dust grains, $a$; and an overall normalization factor that yields the correct mean transit depth.  In addition to these parameters associated with the dust tails, there is also an impact parameter for the transiting planet, and the time of orbital phase zero.  We assume that the hard body of the planet itself does not make a significant contribution to the transit profile.

\begin{figure}
\begin{center}
\includegraphics[width=0.48 \textwidth]{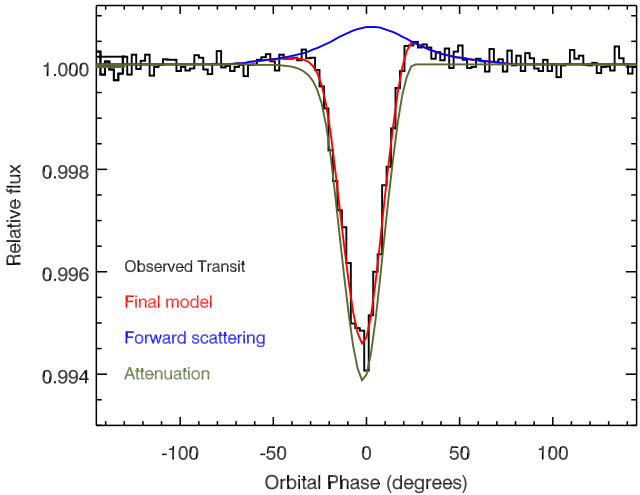}
\caption{Best fitting model transit profile. The black histogram represents the observed transit (see Fig.~\ref{fig:fold}). The red curve is the best fitting model.  The green and blue curves represent the direct absorption and forward scattering components, respectively.  Note how the forward scattering causes the peaks on either side of the transit (see also Budaj 2013).}
\label{fig:model_fit}
\end{center}
\end{figure}

We illustrate in Fig.~\ref{fig:profiles} what some of the possible transit profiles might look like with just two of these parameters varying.  First, we consider only a {\em trailing dust tail}, and we vary the exponential scale length ($\ell$, in units of the host star's radius).  These transit profiles are shown in the top panel of Fig.~\ref{fig:profiles}. The tails' exponential scale length, $\ell$, is color coded to help in distinguishing the different curves.  No forward scattering is included in this case.  Note how the shorter  $\ell$ is, the smaller is the post-transit depression.  The latter becomes quite noticeable when  $\ell$ for the tail becomes greater than about 30\% of the radius of the star.  In the middle panel of Fig.~\ref{fig:profiles} the contribution from forward scattering by relatively large particles ($\sim$$0.5 \,\mu$m) has been included.  In all cases there is a noticeable pre-transit bump due to the forward scattering, but only for the case of short dust tails (i.e., small $\ell$) is there a perceptible post-transit bump; it is typically suppressed by the post-transit depression as the dust tail is in the process of moving off the stellar disk.  The grain sizes assumed here are sufficiently large so that $\lambda/2 \pi a$, the characteristic Mie scattering angle, is comparable with the angular size of the host star as seen from the planet.  In that case, the effective angular scattering pattern from a single grain is  somewhat larger than that of the angular size of the host star (i.e., $\sim$17$^\circ$).  In the bottom panel of Fig.~\ref{fig:profiles} the transit profiles are for the case where there is {\em both a trailing and a leading dust tail}.  The exponential lengths of both dust tails have been taken to be the same (i.e., $\ell_1 = \ell_2$), but the leading tail is taken to have twice the optical thickness everywhere as the trailing tail.  For large $\ell$ there are no pre- or post-transit bumps, as they are pulled below the discernible level by the corresponding depressions caused by the tails.  By contrast, for the case of short tail lengths, i.e., $\ell \lesssim 0.2 \,R_{\rm *}$, there is both a pre- and a post-transit bump, with the latter being slightly larger.  

\subsubsection{Model Fit to the K2-22b Transit}

Finally, we have attempted to fit a simple two-tail dust model to the observed transit profile of K2-22b.  Once the model and its free parameters were selected, we computed the scattering pattern for the dust particles (taken to be of a single fixed size throughout both tails) via a Mie scattering code (Bohren \& Huffman 1983).  Then the scattering pattern was convolved numerically via a 2D integral with the radiation profile coming from the host star (including limb-darkening).  In turn, this net effective scattering profile was convolved in 1D with the assumed exponential tails to produce the scattering pattern as a function of orbital phase.  The attenuation of the beam from the host star was simply computed by placing a double-sided exponential profile in front of a limb-darkened stellar disk in different longitudinal locations (i.e., as a function of orbital phase) and at different vertical locations to get a best estimate of the impact parameter $b$ (in units of the stellar radius).  The sum of the absorption and forward scattering contributions was then added and convolved with the {\em Kepler} long cadence integration time.

We first explored a wide range of parameter space, especially focusing on the physically interesting parameters: $b$, the impact parameter, $\ell_1$, $\ell_2$, $\rho_1/\rho_2$, the exponential scale lengths and ratios of optical thickness of the two tails, respectively, and $a$, the grain size.  We also considered models with a power-law distribution of grain sizes, but these produced no better fits than using a single grain size.  To simplify the search, we fixed $\ell_1 = \ell_2$ under the assumption that the dust sublimation time and the speed away from the planet is similar in the two tails.  However, we allowed $\rho_1/\rho_2$ to be a free parameter since the amount of dust in the two tails could be quite different.  From this broad search, we were able to draw several interesting conclusions.  (1) The best fits, by only a small margin, are found for the single (leading) tail models, i.e., no trailing tail is needed. (2) For single-leading tail models, the impact parameter is constrained to be $0.42 < b < 0.78$ and $0.19 < \ell_2 < 0.48 \,R_*$ (both 90\% confidence limits). (3) For two-sided tail models the allowed ratio of optical thicknesses is constrained to be $\rho_1/\rho_2 < 0.5$ (assuming $\ell_1=\ell_2$).  (4) The best-fitting single grain-size models have $0.3 < a < 0.5\, \mu$m.  The final parameter search and error estimation were done with an MCMC routine. These results are summarized in Table \ref{tbl:planet}.


The overall best-fitting model is for $b = 0.65$ and $\ell_2 = 0.32 \, R_*$.  The results are shown in Fig.~\ref{fig:model_fit}.  The black histogram is the same average transit profile as is shown in Fig.~\ref{fig:fold}.  The red curve is the overall fit to the transit profile, including the convolution with the LC integration time.  The overall model transit profile is comprised of two components: the direct attenuation of the flux from the host star (green curve) and the forward scattering from the dust tail back into the beam directed toward the observer (blue curve). Note how the forward scattering produces small bumps both before and after the transits, with the larger bump following the egress.  

\subsection{Inferred Dust Sublimation Timescale}

The dust sublimation timescale depends on the equilibrium temperature ($T_{\rm eq}$), the mineral composition, and the size of the dust grains (see, e.g., Kimura et al.~2002; Appendix C of Rappaport et al.~2014).  $T_{\rm eq}$ for the dust grains in K2-22b is necessarily somewhat uncertain, ranging between $\sim$1500 K and 2100 K, depending on exactly how it is computed (to within unknown factors of order unity).  However, if we take $T_{\rm eq} = 1700$ K as a typical grain equilibrium temperature, then the lifetime of a 1 micron grain composed of Fe, SiO, or fayalite is a matter of a few tens of seconds.  On the other hand, minerals such as enstatite, forsterite, quartz, and corundum, would have sublimation lifetimes of 1/2, 3, 15, and 30 hours, respectively.  SiC and graphite could last for a very much longer time than any of these.  (All the dust parameters for the above estimates, including the heat of sublimation, are taken from the conveniently tabulated values of van Lieshout et al.~2014). Thus, there is no one ``expected'' dust sublimation timescale that we can anticipate. If on the other hand, we believe that the dust tail is only a few degrees of the orbit in length (see Table \ref{tbl:planet}), and that $\beta \approx 0.05$, then the ``inferred'' lifetime is $\sim 1$ hour (see Eqn.~(4) and (6) of Rappaport et al.~2014 for an explanation), and therefore the dust composition might resemble enstatite or forserite. However, one should not draw too many conclusions from this since the actual tail length depends on a combination of numerous (uncertain) parameters such as the particle sizes (and indeed the size distribution), the parameter $\beta$, the actual equilibrium temperature, and so forth.  Nonetheless, we can say that minerals such as enstatite or forsterite do seem consistent with the observations.

 \subsection{Dust-on-Dust Collisions}

We note that there is an interesting possibility for the dust streaming from the planet to interact with dust that has already been orbiting for a substantial while -- provided that the sublimation lifetime will allow the dust to survive for a sufficiently long time. The time required for orbiting dust in a trailing dust tail (i.e., `outer-track orbits' with $0.05 \gtrsim \beta \gtrsim 0.02$) to meet up again with the planet is approximately $P_{\rm orb}/(2 \beta)$ (see Eqn.~(4) of Rappaport et al.~2014); for K2-22b this amounts to $10-25 \, P_{\rm orb}$ (or $\sim$$4-10$ days).  For particles with $\beta \lesssim 0.02$ the orbits likely lie inside the planet's orbit (`inner-track orbits') with radii between $\sim$94\% and 98\% of $d$.  This leads to synodic orbital periods for those dust particles of between $\sim$10 and 30 $P_{\rm orb}$, fairly similar to the outer-track orbits.  Thus, any orbiting dust that might collide with newly emitted dust must last for $4-10$ days. The dust grains on the inner-track orbits have speeds in the rest frame of the planet that range from $0-6$ km s$^{-1}$ (prograde), while the corresponding range of outer-track orbits is $0-25$ km s$^{-1}$ (retrograde).  Thus, the inner track particles would catch up to, and collide with, newly emitted dust grains with only a few km s$^{-1}$ relative velocity.  By contrast, the outer track particles could collide with newly emitted dust at relatively high speeds of more than 10 km s$^{-1}$.  At such speeds, a collision would deposit a mean energy of $\sim$1 eV per atomic mass unit within the dust grain.  This seems likely to completely destroy most dust particles that happen to collide on the outer track.  

Additionally, any orbiting dust grains which are moving quickly may catch up, and collide, with other more slowly orbiting dust. Again, the relative speeds of any colliding dust particles on the inner track would likely be of order a km s$^{-1}$ and would not necessarily destroy the dust, whereas the reverse is true for the outer-track dust grains.  

We can make a crude estimate of the collision frequency of a dust grain that is orbiting the host star.  Take the mean grain number density along the observer's line of sight during a typical transit to be $n_0$, and the collision cross section to be $\sigma$.  In that case, the collision mean free path is $\ell = 1/(n_0 \sigma)$.  If we take the radial thickness of the dust tail that is causing the transits to be $\Delta r$, then $\Delta r \simeq 1/(n_0 \sigma)$ if both the optical cross section for extinction and the collision cross section are approximately equal (i.e., the geometric cross section appropriate for larger particles), and if the optical depth of the dust cloud during transits is of order unity.  (The latter is necessary since the dust tail is thin in the vertical direction and covers only a small fraction of the disk of the host star (see middle panel of Fig.~\ref{fig:dust_tails}).  Using the above expression, we can estimate the mean collision time, $\tau_{\rm coll}$, for particles orbiting in the dust disk
\begin{equation}
\tau_{\rm coll} \approx \frac{1}{2 \pi} \frac{\Delta r}{d} \frac{v_{\rm orb}}{v_{\rm rel}}\frac{n_0}{\langle n \rangle} \, P_{\rm orb}
\end{equation}
where $v_{\rm orb}$ and $v_{\rm rel}$ are the mean orbital speed in inertial space and the mean relative speeds of the particles in a dust disk, and $\langle n \rangle$ is the mean density in dust grains around the azimuth of the dust disk.  These factors are extremely uncertain, but as an illustrative example, if we take $\Delta r/d \approx 0.05$, $v_{\rm orb}/v_{\rm rel} \approx 10$, and $n_0/\langle n \rangle \approx 100$, then $\tau_{\rm coll}$ is of order several times $P_{\rm orb}$.

What is the fate of dust particles  that collide but are not destroyed in the collision?  There could be locations in the dust disk where the dust is slowed down by the collisions and the density builds up.  In principle, these regions of enhanced density could be approximately fixed in the rotating frame of the planet and host star.  However, since we see only a single transit-like feature during an orbital period, we conclude that such pile-up of dust is not significant anywhere, with the possible exception of near the planet where the dust density is the highest. 

A careful examination of what the effects are of dust-dust collisions in this system or, for that matter, KIC 1255b and KOI 2700b are much beyond the scope of this paper, and we leave that study to another work.

\section{Summary and Conclusions}
\label{sec:concl}

We have reported the discovery in the K2 Field-1 data of a new planet that is likely the third example of a planet disintegrating via the emission of dusty effluents.  The planet is in an ultrashort 9.1457 hour orbit about an M star.  The evidence we presented for the presence of a dust tail includes erratically and highly variable transit depths ranging from $\lesssim 0.14\%$ to 1.3\% in both the K2 data as well as in follow-up ground-based observations from five different observatories.  The folded orbital light curve from the K2 data exhibits a clear post-egress `bump' and a much less significant, but plausible, pre-ingress `bump', that are not found in conventional hard-body transits.  There is no post-egress depression in the flux as was seen in KIC 1255b or in KOI 2700b.  On at least one observation with the GTC, the transit depths were distinctly wavelength dependent with the transits $\sim$25\% shallower at 840 nm than at 630 nm.  While this goes in the same direction as limb-darkening effects (see, e.g., Knutson et al.~2007; Claret \& Bloemen 2011), we argue that the magnitude of the limb-darkening effect explains only a small fraction of what is observed.  Furthermore, the variable behavior of the color-dependent transit depths strongly suggests an origin in dust scattering.  This requires relatively non-steep power-law particle size distributions with $\Gamma \simeq 1$ to $3$ with maximum sizes in the range of $0.4-0.7$ $\mu$m}.

The host star is an M star with $T_{\rm eff} \simeq 3800$ K.  There is a companion star 1.9$''$ away with $T_{\rm eff} \simeq 3300$ K.  We infer a distance to the system of $225 \pm 50$ pc, which, in turn, implies a projected physical separation of the two stars of $\sim$430 AU.  At this distance, the companion star may have been incidental to the formation of the planet, but perhaps was instrumental in driving it toward the host star via Kozai-Lidov cycles with tidal friction (Kozai 1962, Lidov 1962;  Kiseleva et al.~1998; Fabrycky \& Tremaine 2007).  

The minimum transit depth of $\lesssim 0.14\%$ sets an upper limit to the size of the underlying hard-body planet of 2.5 $R_\oplus$ for an assumed radius of the host star of $0.57 \pm 0.06 \,R_\odot$.  This is consistent with the low surface gravity that is required to drive off metal vapors that could condense into dust (Rappaport et al.~2012; Perez-Becker \& Chiang 2013).  In fact, it seems likely if the dust-emitting scenario we report for K2-22b is correct then Mars, Mercury, or even lunar sized bodies with surface gravities of 1/6 to 1/3 that of Earth are to be preferred.  

We find that the dust tail of K2-22b  has two properties that are distinct from those of KIC 1255b\footnote{We note that Bochinski et al.~(2015) were able to measure a small difference in transit depths in KIC 1255b between g' and z' bands, and thereby concluded that the dusty effluents in this object contained a component of larger grains in the range of $0.25-1 \, \mu$m.}  or KOI 2700b: (1) a leading dust tail (as opposed to a trailing one); and (2) a characteristic (e.g., exponential) scale length for the dust tail that is $\lesssim 1/2 \,R_{\rm *}$.  The former requires dust transported to about twice the planet's radius in the direction of the host star until it effectively overflows its Roche lobe, thereby going into an orbit that is faster than that of the planet.  It is also necessary to have $\beta$ (the ratio of radiation pressure forces to gravity) be $\lesssim 0.02$ which would be the case for very low luminosity host stars in combination with very small ($\lesssim 0.1\,\mu$m) or very large ($\gtrsim 1 \, \mu$m) dust particles (see Fig.~\ref{fig:betas}).  The shorter dust tails could result from a combination of very low values of $\beta$ and a short dust-grain sublimation time of $\lesssim$ an hour.  

Mass loss rates in the form of high-Z material from this planet are likely to be $\approx 1.5 \times 10^{11}$ g s$^{-1}$ implying a lifetime of $20-70$ Myr, for planet masses in the range of the moon's to Mercury's mass.

The scenario described above for explaining the various features of the transit light curve and its variability is not entirely self-consistent.  The requirement for particles to enter a leading tail is $\beta \lesssim 0.02 \, \mu$m which implies particle sizes of either $a \lesssim 0.1 \,\mu$m or $a \gtrsim 1 \, \mu$m.  At the same time, the post-transit bump requires a forward scattering peak that is comparable to the angular size of the host star.  This corresponds to particle sizes of $\sim 1/2 \, \mu$m.  Finally, the color dependence of the transits observed with the GTC implies a non-steep power-law size distribution with a maximum size of $\sim$$1/2 \,\mu$m.  Thus, larger particles could account for all three of these observational pieces of evidence---but only if a substantial fraction of the particles are large (e.g., $\sim 1 \,\mu$m).  This requirement for such large particles seems somewhat unusual in comparison with other known astrophysical collections of dust such as in the ISM (e.g., Mathis et al.~1977; Bierman \& Harwitt 1980), Solar-system comet tails (e.g., Kelley et al.~2013), Io (e.g., Kr\"{u}ger et al.~2003a; 2003b), and the Earth's atmosphere (e.g., Liou 2002; Holben et al.~1998).

The host star K2-22 shows photometric modulations with an amplitude of approximately 1\% and a timescale of 15 days. It is interesting to note that the other two host stars to disintegrating planets also display photometric modulations at the several percent level, likely associated with star spots. It has been suggested that stellar activity may play an important role in modulating the process that generates the dust (Kawahara et al.~2013; Croll et al.~2015). The large spot modulations for the host stars of the candidate disintegrating planets is suggestive and warrants further investigation. 

Finally, we summarize in Table \ref{tbl:dusty} some comparative properties of the three known `disintegrating planets'.  Hopefully, further patterns of similarity will emerge as more of these objects are discovered.

\begin{deluxetable}{lccc}

\tablecaption{Comparison of Dusty Planets \label{tbl:dusty}}
\tablewidth{0pt}

\tablehead{
\colhead{Parameter (units)} &
\colhead{EPIC 201637\tablenotemark{1}} &
\colhead{KIC 1255b\tablenotemark{2}} &
\colhead{KOI 2700b\tablenotemark{3}} 
}

\startdata
$P_{\rm orb}$ (hr) & 9.146 & 15.68 & 21.84  \\
Depth (\%) & $0-1.3$ & $0-1.4$  & $0.031-0.053$ \\
Variability & highly &  highly  & slowly  \\
$d/R_*$ & $3.3 \pm 0.2$ &  $4.3 \pm 0.4$ &  $5.9 \pm 0.4$ \\
$\theta_*$ (deg) & 17 &  13 &  10  \\
Pre-bump & weak &  yes &  ... \\
Post-bump & yes &  no &  no \\
$\theta_{\rm tail}$ (deg)\tablenotemark{4} & $\lesssim 8$ & $\sim$$10-15$  & $\sim$24  \\
$\theta_{\rm tail}/\theta_*$ & $\lesssim 0.5$ & $\sim$$0.77-1.1$  & $\sim$2.4  \\
$T_{\rm host}$ (K) & 3830 &  4300 &  4435  \\
$T_{\rm eq}$ (K)\tablenotemark{5}  & 2100 &  2100 	&  1850  \\
$\beta_{\rm max}$ & 0.05  & 0.1  & 0.07  \\
$\dot M_{\rm dust}$ ($10^{10}$ g/s) & 20 & 20 & 0.15  
\enddata
\tablenotetext{1}{Results from this work.}
\tablenotetext{2}{Values from Rappaport et al.~(2012); Brogi et al.~(2012); Budaj (2013); van Werkhoven et al.~(2014).}
\tablenotetext{3}{Values from Rappaport et al.~(2014).}
\tablenotetext{4}{Approximate exponential tail length in degrees of orbital phase.}
\tablenotetext{5}{$T_{\rm eq} \equiv T_{\rm eff} \sqrt{R_*/d}$. }

\end{deluxetable}


Discoveries of new disintegrating planets in the upcoming K2 fields could be potentially quite important. If found, these candidates are likely to orbit brighter stars than the host stars of the examples discovered to date. It would also be interesting to discover them orbiting a richer variety of host stars, especially since we have inferred from K2-22b that the low luminosity of the host star plays an important role in determining the trajectories of the dust flowing from these very special planets. Future follow-up observations with ground-based telescopes, both in photometric and spectroscopic modes, are likely to provide us with further information about the process that generates the dust emission, and give us better insights into the relevant physical processes in these extreme-environment systems.

\acknowledgements We thank Joshua Pepper and Smadar Naoz for helpful comments  and an anonymous referee for suggestions that greatly improved the presentation of the material. We thank Allyson Bieryla and Dave Latham for helping us with the FLWO observations of the object. We are grateful to Evan Sinukoff, Erik Petigura, and Ian Crossfield for observing this target with Keck/HIRES. We thank Yi Yang for helping with the HSC images, and Takuya Suenaga for volunteering his time to do so. The help from the Subaru Telescope staff is greatly appreciated. We acknowledge Tsuguru Ryu for his support on the Okayama transit observations. T.H. is supported by Japan Society for Promotion of Science (JSPS) Fellowship for Research (No. 25-3183). I.R. acknowledges support from the Spanish Ministry of Economy and Competitiveness (MINECO) and the Fondo Europeo de Desarrollo Regional (FEDER) through grant ESP2013-48391-C4-1-R. We extend special thanks to those of Hawaiian ancestry on whose sacred mountain of Mauna Kea we are privileged to be guests.  Without their generous hospitality, the Keck observations presented herein would not have been possible. This work was performed, in part, under contract with the Jet Propulsion Laboratory (JPL) funded by NASA through the Sagan Fellowship Program executed by the NASA Exoplanet Science Institute. This article is partly based on observations made with the Nordic Optical Telescope operated by the Nordic Optical Telescope Scientific Association and the Gran Telescopio Canarias operated on the island of La Palma by the IAC at the Spanish Observatorio del Roque de los Muchachos. This research has been supported by the Spanish MINECO grant number ESP2013-48391-C4-2-R.  The Infrared Telescope Facility is operated by the University of Hawaii under contract NNH14CK55B with the National Aeronautics and Space Administration.  FM acknowledges the support of the French Agence Nationale de la Recherche (ANR), under the program ANR-12-BS05-0012 Exo-atmos.

\end{document}